\documentclass[10pt,twocolumn]{interact}

\usepackage{graphicx}
\usepackage{stfloats}
\usepackage{midfloats}
\usepackage{array}
\usepackage{booktabs}
\usepackage{multirow}
\usepackage{verbatim}
\usepackage{cite}
\usepackage{flushend}
\usepackage{float}
\newcommand{\head}[1]{\textnormal{\textbf{#1}}}

\usepackage{color}

\emergencystretch=\maxdimen
\usepackage{epstopdf}
\usepackage{subfigure}

\usepackage[numbers,sort&compress]{natbib}
\bibpunct[, ]{[}{]}{,}{n}{,}{,}
\makeatletter
\def\NAT@def@citea{\def\@citea{\NAT@separator}}
\makeatother

\theoremstyle{plain}

\theoremstyle{definition}

\theoremstyle{remark}

\begin{document}

\begin{onecolumn}

\title{Fast full-body reconstruction for a functional human RPC-PET imaging system using list-mode simulated data and its applicability to radiation oncology and radiology}

\author{
\name{Paulo~Magalhaes~Martins\textsuperscript{a}\textsuperscript{b}\textsuperscript{c}\thanks{CONTACT
    Paulo~Magalhaes~Martins. Email: p.martins@dkfz.de; Address: Department of Medical Physics in
  Radiooncology, German Cancer Research Center - DKFZ, Heidelberg,
Germany; Paulo Crespo. Email: crespo@lip.pt; Address: LIP - Laborat\'orio de
  Instrumenta\c{c}\~ao e	F\'isica Experimental de Part\'iculas, Physics Department, University of
  Coimbra, Coimbra, Portugal.}, Paulo Crespo\textsuperscript{b}\textsuperscript{d},
Miguel Couceiro\textsuperscript{b}\textsuperscript{e}, Nuno
Chichorro Ferreira\textsuperscript{f}\textsuperscript{g}, Rui Ferreira
Marques\textsuperscript{b}\textsuperscript{d}, Joao Seco\textsuperscript{a}\textsuperscript{h} and
Paulo Fonte\textsuperscript{b}\textsuperscript{e}}
\affil{\textsuperscript{a}Department of Medical Physics in
  Radiooncology, German Cancer Research Center - DKFZ, Heidelberg,
Germany; \textsuperscript{b}LIP - Laborat\'orio de
  Instrumenta\c{c}\~ao e	F\'isica Experimental de Part\'iculas, Physics Department, University of
  Coimbra, Coimbra, Portugal; \textsuperscript{c}Institute of Biophysics and Biomedical
Engineering - IBEB, Faculty of Sciences of the University of Lisbon,
Lisbon, Portugal; \textsuperscript{d}Physics Department,
  University of Coimbra, Coimbra,
  Portugal; \textsuperscript{e}Polytechnic Institute of Coimbra, ISEC,
  Coimbra, Portugal; \textsuperscript{f}IBILI - Institute for
  Biomedical Imaging
  and Life Sciences, Faculty of Medicince of the University of
  Coimbra, Coimbra, Portugal; \textsuperscript{g}ICNAS - Instituto de Ci\^encias Nucleares Aplicadas  \`a
  Sa\'ude, University of Coimbra, Coimbra,
  Portugal; \textsuperscript{h}Department of Physics and Astronomy,
  University of Heidelberg, Heidelberg, Germany}
}

\maketitle

\begin{abstract}
{\bf Background:} Single-bed whole-body positron emission tomography based
on resistive plate chamber detectors (RPC-PET) has been proposed for
human studies, as a complementary resource to scintillator-based PET
scanners. The purpose of this work is mainly about providing a
reconstruction solution to such whole-body single-bed data collection
on an event-by-event basis. We demonstrate a fully three-dimensional
time-of-flight (TOF)-based reconstruction algorithm that is capable of
processing the highly inclined lines of response acquired from a
system with a very large axial field of view, such as those used in
RPC-PET. Such algorithm must be sufficiently fast that it will not
compromise the clinical workflow of an RPC-PET system.

\noindent {\bf Material and methods:} We present simulation results from a voxelized
version of the anthropomorphic NURBS-based cardiac-torso (NCAT)
phantom, with oncological lesions introduced into critical regions
within the human body. The list-mode data was reconstructed with a
TOF-weighted maximum-likelihood expectation maximization (MLEM). To
accelerate the reconstruction time of the algorithm, a multi-threaded
approach supported by graphical processing units (GPUs) was
developed. Additionally, a TOF-assisted data division method is
suggested that allows the data from nine body regions to be
reconstructed independently and much more rapidly.
 
\noindent {\bf Results and Conclusions:} The application of a TOF-based scatter
rejection method reduces the overall body scatter from 57.1\% to
32.9\%. The results also show that a 300-ps FWHM RPC-PET scanner
allows for the production of a reconstructed image in 3.5 minutes
following a 7-minute acquisition upon the injection of 2 mCi of
activity (146 M coincidence events). We present for the first time a
full realistic reconstruction of a whole body, long axial coverage,
RPC-PET scanner. We demonstrate clinically relevant reconstruction
times comparable (or lower) to the patient acquisition times on both
multi-threaded CPU and GPU.

\end{abstract}

\begin{keywords}
RPC-PET; time-of-flight; parallel computing; lesion detectability
\end{keywords}

\clearpage

\end{onecolumn}

\oddsidemargin        14.32mm
\evensidemargin       14.32mm
\addtolength{\oddsidemargin}{-1in}
\addtolength{\evensidemargin}{-1in}
\columnsep         1pc
\textwidth        43pc

\begin{twocolumn}

\section{Background}

Single-bed whole-body positron emission tomography based on
resistive plate chamber detectors~\cite{FonteNIMA2000}
has been proposed for human
studies~(RPC-PET)~\cite{BlancoNIMA2003}.
RPC-based detectors offer simple and economical construction,
reliability of operation, and
extremely good time
and intrinsic position resolutions
(300~ps full width at half maximum (FWHM) for the coincidence of two
511 keV annihilation photons~\cite{BlancoNIMA2003}, and
0.4~mm FWHM for a small-animal RPC-PET prototype~\cite{Martins2014}, respectively).
The depth-of-interaction (DOI) can also be accurately
measured~\cite{BlancoTNS2006}, rendering RPC-PET essentially
parallax-free. The lack of energy resolution in the RPC-PET
detectors is compensated for by their very-high time-of-flight (TOF)
resolution and their energy sensitivity~\cite{Blanco2009}.
These properties make RPC detectors suitable for large axial field-of-view (AFOV)
TOF-PET system.

Fig.~\ref{fig:rpcpet} illustrates a planned RPC-PET system integrated
with a computed tomography (CT) machine. In addition to providing
multimodality imaging, the CT will also provide the data necessary to
estimate the attenuated and object-scattered events (the latter by
simulations~\cite{Werner2006,Watson2007}) and introduce those corrections into the
reconstruction.

\begin{figure}[!h]
\vspace{0mm}
\centering
\includegraphics[width=3.2in]{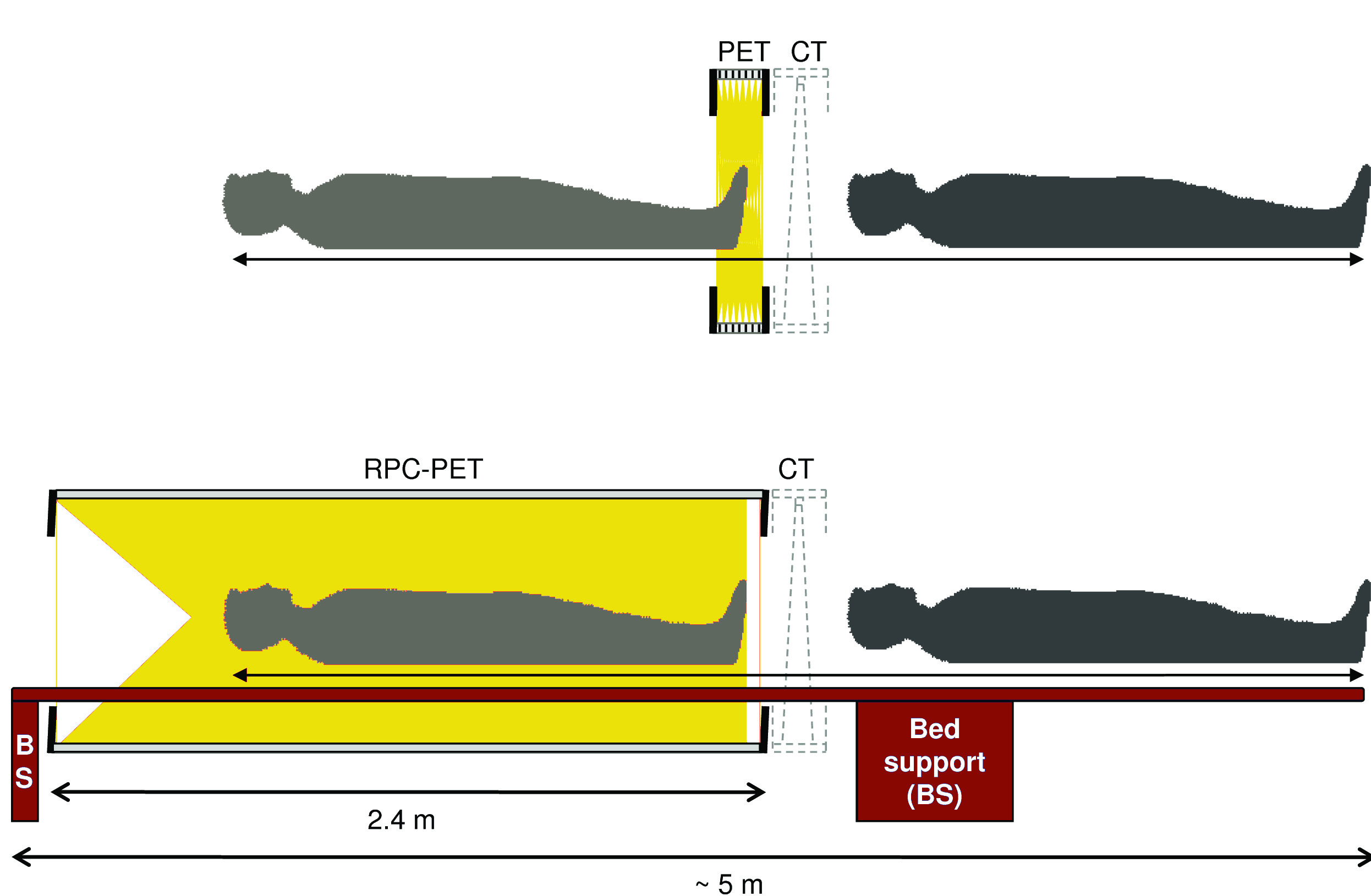}
\caption{Scheme of the planned RPC-PET system
integrated with a CT machine. Detailed transverse and
longitudinal views of this RPC-PET system and its detector layout
have been presented by Couceiro et al.~\cite{CouceiroNIMA2012}.
The full system length is well within the typical dimensions of current PET-CT rooms
utilized for clinical scans. The PET-CT center of the
University of Coimbra, for example, has two rooms
each with a total length greater than 7\,m.
}
\label{fig:rpcpet}
\end{figure}

In the last decade, several authors have studied the feasibility of extended-AFOV crystal-based
PET scanners\mbox{~\cite{Crosetto2003,Eriksson2008,Borasi2010,Poon2012}}.
However, the high cost of such 
systems and the slow performance of 
reconstruction
algorithms remains a major concern.

In this work we address the challenge of
providing a reconstruction routine tailored to the potentialities of
RPC-PET. Such a routine must process data from the whole AFOV of the
system, rather than from one bed at a time as is the case with
scintillator-based tomographs with shorter AFOV. In addition, the same routine must be able to
incorporate the TOF advantage provided by RPC detectors. At the
same time, it must also be fast enough to manage the imaging speed
that an RPC-PET system may provide due to its increased sensitivity,
which would not only be used to scan faster. It could be used to reduce the injected dose or to have
shorter frames in dynamical acquisitions.

The first two aforementioned reconstruction challenges, incoming
data from the whole AFOV and capability of TOF processing, 
were approached by choosing an iterative algorithm capable of
processing data in list-mode format: the TOF-weighted 
maximum-likelihood expectation maximization (MLEM)~\cite{Shepp1982, Groiselle2004}. 
It was coupled to an attenuation correction procedure developed 
to accept events from the whole patient body. It is
widely accepted today the combination of list-mode acquisition and iterative reconstruction of
3D data, as list-mode storage
is more efficient than the binned format for 3D data due to the large
number of measured LORs with respect to the number of detected
events~\cite{karpJNM2008}. List-mode reconstruction methods have greatly
improved with the inclusion of both TOF and other
physical effects in the system
model~\cite{Crespo2007,Witherspoon2012,Conti2013}. The
highly-computational demands of list-mode reconstruction for shorter
AFOV systems has been
successfully overcome with fast computers and parallel computing
methods demonstrating very impressive reconstruction
times~\cite{Pratx2009,Pratx2011,Cui2011,Sportelli2013}. With this work
we demonstrate a remarkable reconstruction time for large AFOV
systems.

\section{Materials and Methods}

The reconstructed images presented in this article resulted from the
application of the RPC-PET reconstruction algorithm to GEANT4
simulated data (versions 9.02 and 9.03). The software-based
anthropomorphic NCAT phantom~\cite{Segars2001}
was adapted to Geant4, including the whole body activity distribution and
attenuated photon emissions from the human body, taking into consideration
different tissues and densities~\cite{Crespo2012}. 
This reconstruction was processed with
self-designed C, C++ and CUDA routines. The phantom used in the
reconstruction was placed on a
body-centered parallelepiped volume and final images were reconstructed
into a 350~$\times$~350~$\times$~1000 matrix with
2~mm~$\times$~2~mm~$\times$~2~mm voxels. The
analyzed list-mode outputs were constructed for an RPC-PET system with
120 gaps and a singles detection efficiency to 511~keV perpendicular gamma rays of
19.4\%~\cite{Crespo2012}. The output images were analyzed
and processed with ROOT (version 5.34.00, CERN, Geneva, Switzerland).

\subsection{Simulated activity}

\subsubsection{Six spheres in a homogeneous background}

This study considered
six simulated spherical sources placed in a homogeneous background, with a
signal-to-background activity ratio of 6:1. The diameter of each
sphere was 10~mm with their locations forming a hexagon with 80~mm
sides. The phantom was 1.1~m long 
and had a diameter of 35~cm (volume = 105~L). Assuming
fluorodeoxyglucose ($^{18}$F-FDG) as the
decaying solution and a typical body background activity
concentration of 2.12~kBq/mL~\cite{Crespo2012},
this corresponded to a scan time of 88~s and 20~billion decays. From 142~M detected events,
41~M were true
events and the scatter fraction (SF) after the scatter rejection
by a 300~ps FWHM TOF resolution was 45.2\%, which was explained by the large
diameter of the cylinder that in turn corresponded to a patient with a large
body mass index (BMI).  The axial acceptance angle was restricted to
less than $45^{\circ}$, since the gain in sensitivity does not compensate the
increase in the SF; this left 36~M true events to be reconstructed.

\subsubsection{NCAT anthropomorphic phantom}

Typical standard uptake values (SUV) in tissue range between 0.2 and 0.8~g/mL,
with this variability being ascribed to variables such as 
patient weight, blood glucose level, length of uptake period,
partial-volume effect, recovery coefficient, and type of region of interest
(ROI)~\cite{Paquet2004}. In this simulation, we assumed a tissue SUV of
0.27~g/mL which, considering a patient weight of 92kg, yielded a tissue 
activity concentration of 0.21~kBq/mL for an injected activity of
2~mCi. After the typical 1~h waiting time for
$^{18}$F-FDG uptake, approximately 1~mCi of the administered activity was
retained~\cite{Swanson1990}.
To study the NCAT anthropomorphic phantom, 1.6 $\times 10^{10}$ decays were
considered, which corresponded to an acquisition time of 440~s. 
The RPC-PET system detected
146~M coincidence events. 
The axial acceptance angle was restricted to
less than $45^{\circ}$ and the NCAT phantom was displaced by 256 mm in the
direction from head to feet in order to increase the sensitivity in
the upper region of the body. After rejecting
the scattered events by a 300~ps FWHM TOF resolution, the SF
was 32.9\% for a total of 49~M true events.

\begin{figure}[!b]
\centering
\includegraphics*[width=8cm]{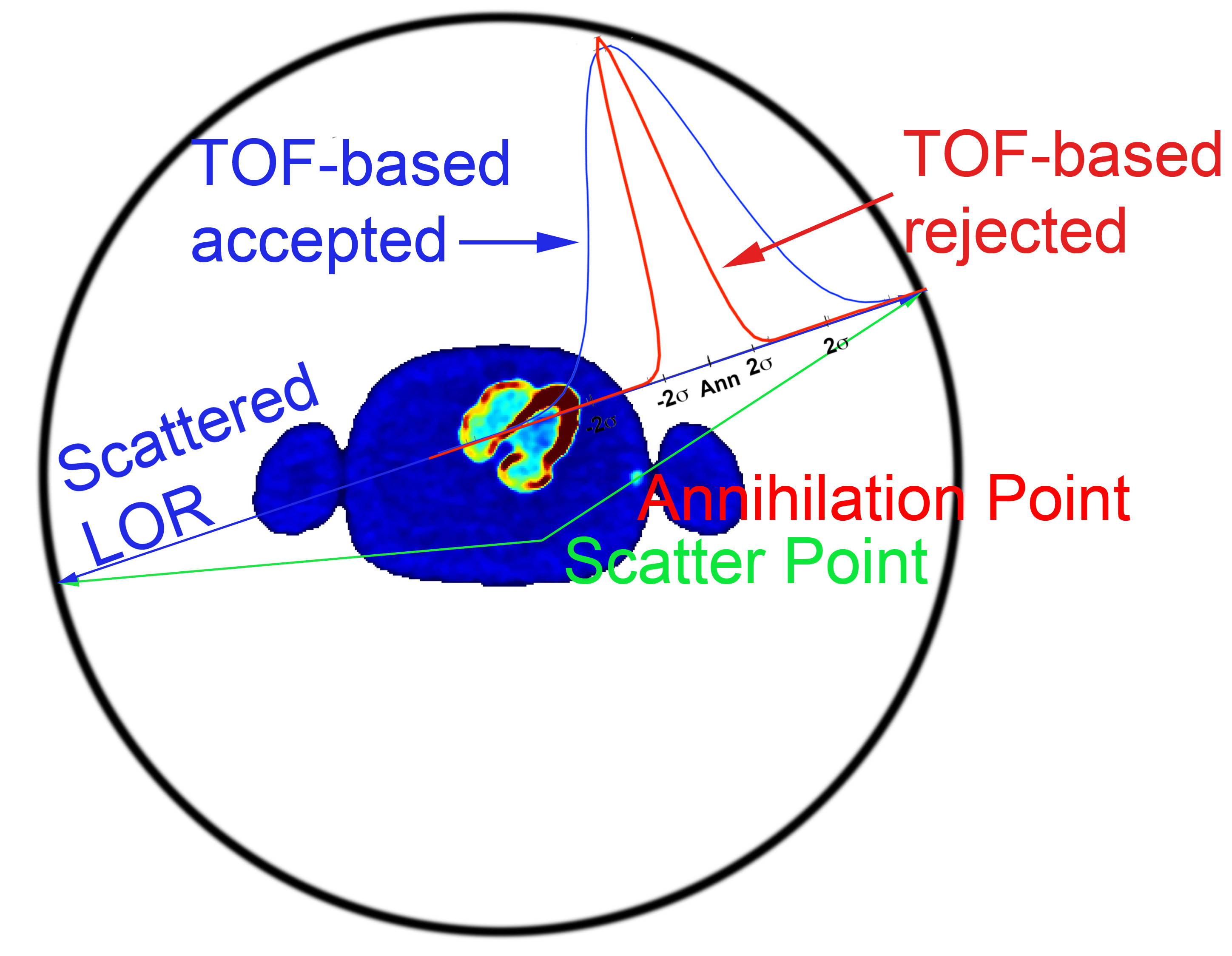}
\caption{Scheme of the scatter rejection method. The calculated
  annihilation point located in the scatter LOR has a certain
  probability of being inside the human body, depending on the TOF
  resolution. Poorer TOF resolutions (curve in blue) increase the probability of the
  annihilation point being within the human body, resulting in a
  TOF-based accepted event. Better TOF resolutions reject a higher
  fraction of scattered events. The limits of rejection were
  estimated at the 95\% confidence level.}
\label{fig:SRF_draw}
\end{figure}

\subsubsection{Six lesions in an anthropomorphic phantom}

Datasets of lesions with a diameter of 10~mm each were
inserted by simulation into the anthropomorphic phantom with a lesion to
background activity ratio of 10:1~\cite{Zhang2007}
and placed in the following regions: cervical, sub-clavicular,
axillary, inguinal, knee and foot~\cite{Zhang2007}. To determine the
events to be simulated in each lesion and detected by the RPC-PET
system, we calculated the time-integrated activity density of the
background phantom tissue. This tissue contributed 
60\% of the total body activity and occupied a volume of 82~L; therefore,
it had a time-integrated activity density of
4.4~cts~mm$^{-3}$. Consequently, 20.8~k events were simulated for each
lesion.

\subsection{TOF-based scatter rejection of anthropomorphic events}
\label{sec:SRF_method}

A method to increase lesion detectability based on the rejection of
scattered events by means of their TOF information was
investigated.

Fig.~\ref{fig:SRF_draw} illustrates this method
by which the annihilation point determined along the scatter
line-of-response (LOR)
had a certain probability of being inside the human body, according
to the TOF resolution. The limits of rejection were
estimated at a 95\% confidence level. The body outline was determined
from knowledge of the NCAT outline and a CT would be used for measured
data.

\subsection{Multi-threaded GPU-based parallelization reconstruction
  strategy with a TOF kernel}
\label{sec:tof_kernel}

In this reconstruction, the TOF kernel was modeled as a Gaussian with
a kernel of 300~ps FWHM - measured coincidence time resolution,
as shown by experimental results with an RPC-PET detector for human PET~\cite{BlancoNIMA2003}. This modeling was
based on the direct technique for voxel filling~\cite{Crespo2007}.

The reconstruction routine consisted of considering a certain number of samples
equally spaced inside the 300~ps FWHM TOF Gaussian and giving them the
equivalent Gaussian weight, so the sum of the weights equaled
one.  These samples are
centered in the coincidence time difference of the corresponding two
annihilation photons detected. Fig.~\ref{fig:gaussian_dots} shows the equally spaced samples with
their corresponding weights.

\begin{figure}[!h]
\centering
\includegraphics*[width=5cm]{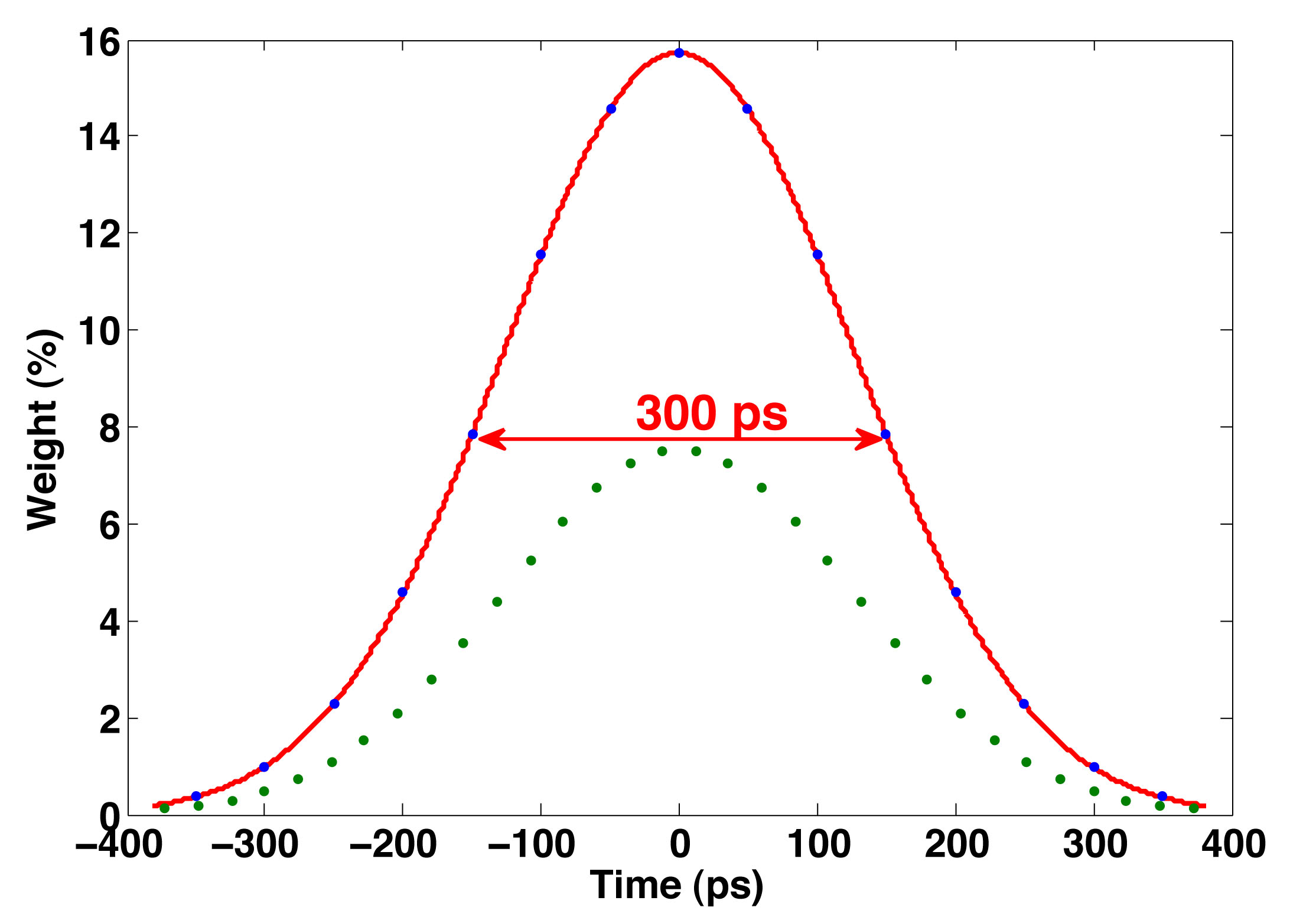}
\vspace{-2mm}
\caption{Distribution of the time uncertainty associated with a 300~ps
  FWHM TOF resolution. The TOF uncertainty along the space variable and its respective
  weight is distributed along 15 and 32 samples, used for the 16-thread CPU
  and GPU routine implementations, respectively. These values
  are accessed through a look-up table (LUT).}
\label{fig:gaussian_dots}
\end{figure}

The more samples we considered, the lower
was their individual weight. For the 16-thread CPU implementation, 15 samples were
used while for the GPU implementation, 32 samples were considered
without compromising reconstruction speed and image quality.
In order to increase
the compute-to-memory access ratio, we used the GPU constant memory. To overcome the
inefficiency of cache entries resulting from
having the TOF uncertainties along the space variable and their corresponding weights stored in separate
arrays, we arranged the elements of these arrays as a struct,
known as an array of structs~\cite{Hwu2010, Sanders2011}.

In a first approach, we gave an equal weight to each sample of
the TOF-kernel and this was filled randomly~\cite{Martins2014b}, thus
consuming undesired computing time capabilities. In the following results, we accessed the
300~ps FWHM Gaussian samples through a look-up table (LUT) containing
the TOF uncertainties along the space variable and their respective
weights. This approach always generated an identically reconstructed
image, in contrast with the random sample generation method. Our
studies revealed that the former has increased background noise and
the latter serves as a smoothing filter with a small impact on
contrast recovery.

The reconstruction routines were both implemented on an NVIDIA Tesla
C2075 GPU assisted by a double Intel Xeon E5620 2.4GHz CPU with 16
threads, versus the 16-thread CPUs alone.

\subsection{MLEM and OSEM mathematical implementation}

The list-mode data were reconstructed by using MLEM that included a TOF probability density function~\cite{Crespo2007}.
Attenuation correction was implemented in the forward projection by a
weight inversely proportional to the probability of the event being
attenuated. The
attenuation correction was incorporated neither 
in the form of the attenuation weighted (AW)-MLEM~\cite{Hebert1990}
nor in the AW-OSEM~\cite{Comtat1998}. It has been shown that implementing the attenuation
correction to the LOR is a good approximation for applying a
correction in the sensitive
matrix~\cite{Levkovitz2001,Groiselle2004}. The inclusion of a random
and TOF-based scatter corrections remained outside the scope of this
work.

Sieves~\cite{Snyder1985} (an operation that suppresses high
frequency noise) can be applied to the image estimate after each
iteration to impose smoothness and reconstruction stability. For this
reason, we introduced a median filter between MLEM iterations in our
reconstructions routines. In fact, this operation was fundamental for
the reconstruction of very low data for such a huge volume. The
drawback of this technique is the impact on the reconstruction speed
which still remains its bottleneck despite the parallelization of the
image regions to be filtered.

\subsection{Performance optimization: TOF-assisted data division into different body regions}
\label{sec:data_division_method}

This method proposed to increase the performance of the reconstruction
routine by independently reconstructing nine different
regions of the body. 

In Fig.~\ref{fig:division_scheme}, we see that
for LOR 1 the most probable annihilation location given by a TOF-based
calculation ascribed a 40\% probability of such an event occurring within
image 5, and a 60\% probability of such an event having arisen in image
region number 6.
\begin{figure}[!h]
\centering
\includegraphics*[width=.4\textwidth]{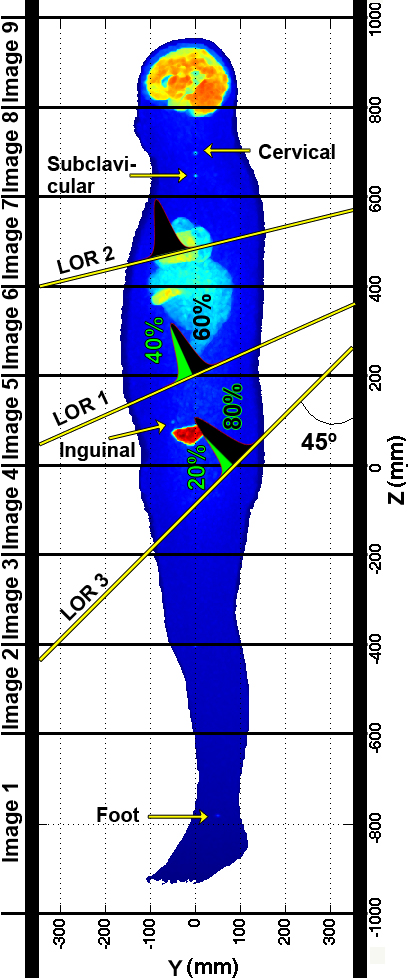}
\vspace{-3mm}
\caption{TOF-assisted data division into nine different body
  regions. This maximum intensity projection (MIP) image clearly shows
the inguinal, subclavicular, cervical, and foot lesions. The
separation between the inguinal lesion and the bladder is remarkable,
as discussed later in the text.}
\label{fig:division_scheme}
\end{figure}
However, LOR 2 was assigned in its entirety to image region number 7. Even
very-inclined LORs (45$^{\circ}$) crossing several image spaces, such as LOR 3,
always contributed to 1 or 2 image regions provided that the TOF
information remained much smaller than the width of each region. 

Data
incoming from an outer shell of 6~mm in both axial directions to these
image regions was added and an overlap between neighbouring regions
was performed as is the usual practice in conventional multi-bed
systems.

To consider the
events laying outside the divided images while iterating on the TOF-kernel, a margin
of $\approx$~3$\sigma$ in the space variable was taken into consideration.
In this method, the first image region (feet) is 200 $\times$ 350 $\times$ 350. The
other 8 image regions are 100 $\times$ 350 $\times$ 350. Image regions 2 to 8
are extended to 176 $\times$ 350 $\times$ 350 ($\sigma = 19.11$ mm $\Rightarrow$ 76 mm =
3.77 $\sigma$) to include the tails of the Gaussian TOF kernel filling in
both axial directions. Image region number 1 is extended to 240 x 350
x 350 in the upside direction (4.19 $\sigma$). Image region number 9 is
extended to 144 $\times$ 350 $\times$ 350 in the downside direction
(4.19 $\sigma$).   

The nine different images were then summed, resulting in a single 
volume similar to the whole-body reconstructed image.

\vspace{-0.5cm}

\subsection{Contrast recovery coefficient calculation}

Following the strategy adopted by~\cite{Witherspoon2012} to determine the local
contrast recovery coefficient (CRC) values, independent
reconstructions were performed both on the NCAT simulated data without
the inserted lesions assumed as background, and after inserting the
lesions by simulation, the latter representing the signal. The Volumes of Interest
(VOIs) were drawn for each lesion with the respective diameter on both
images and the CRC was calculated as

\begin{equation}
CRC = (H/B -1)/(a-1),
\end{equation}
where {\bf{\emph{H}}} is the average intensity in the lesion VOI,
{\bf{\emph{B}}} is the average intensity in the corresponding
background VOI, and {\bf{\emph{a}}} is the simulated lesion to background activity
ratio of 10:1. The CRC values were calculated for a single
realization and their uncertainties were derived from the uncertainties in the
region average of the signal and the background~\cite{Tong2010}. The same procedure was applied to the six spheres in the
homogeneous background. In this case, the simulated contrast was 6:1.

\section{Results}

\subsection{Six spheres in homogeneous background}

\begin{figure}[!t]
\centering
\includegraphics[width=3.in]{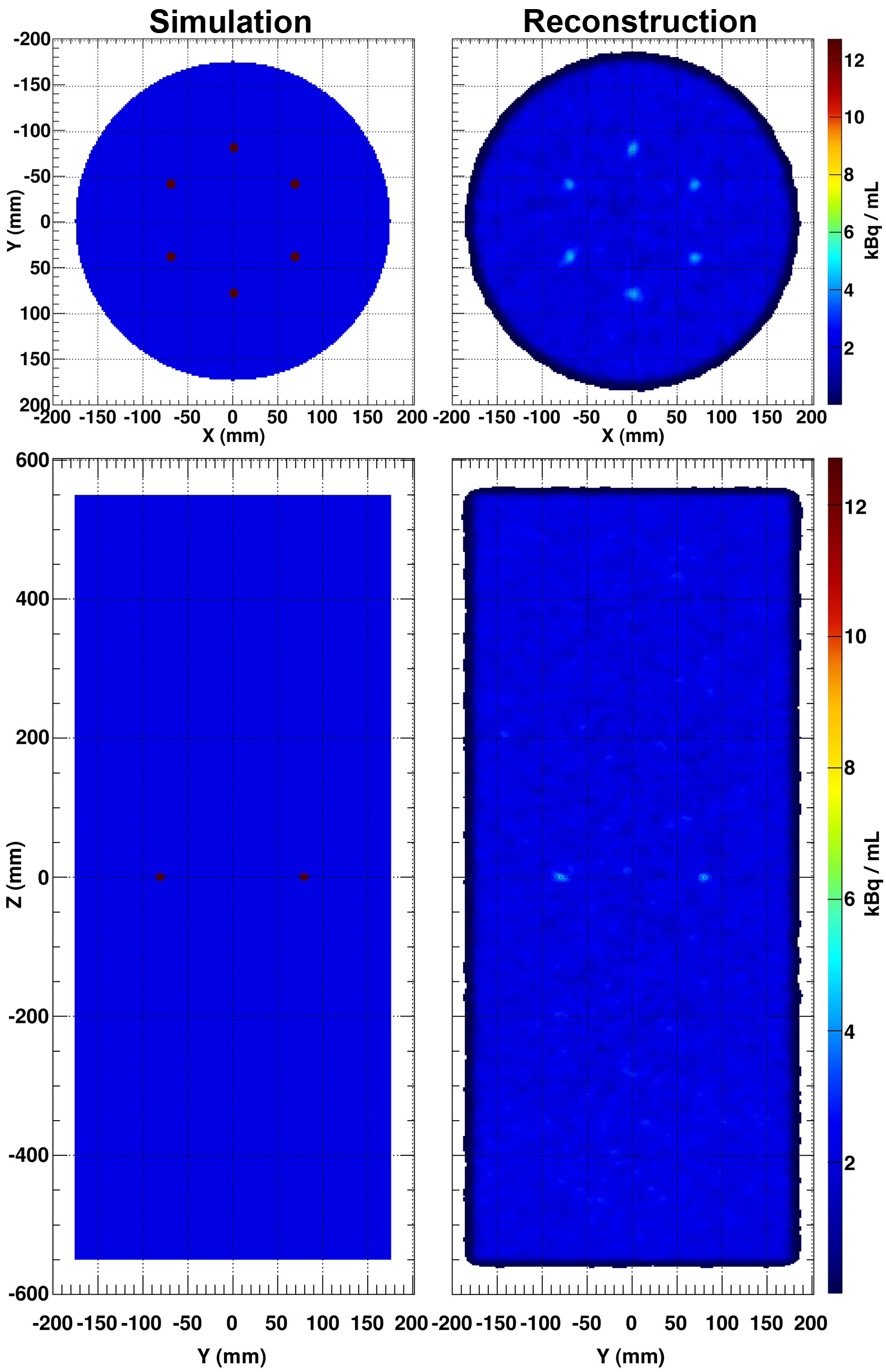}
\vspace{-2mm}
\caption{Six simulated and reconstructed spherical sources immersed in
  a homogeneous activity background. The top images represent the
  axial view of the cylinder with a 35~cm diameter:
  simulated emission phantom (left); reconstructed image (right). The bottom row presents the
  corresponding sagittal views. A
  cut on the intensity scale was performed on the
  reconstructed images. The reconstructed image effectively reproduces the simulation
  phantom, despite the lower visual contrast of the spheres.}
\label{fig:6spheres}
\end{figure}

Fig.~\ref{fig:6spheres} shows six simulated and reconstructed -- after
20 MLEM iterations -- spherical sources
immersed in a homogeneous background with a signal-to-background
activity ratio of 6:1.
The reconstruction and simulated phantom show 
good quantitative agreement between background values despite the
lower contrast of the spheres in the reconstructed images and the increased
statistical noise and background variability caused by a low
statistics dataset.

\subsection{Reconstructed images}

Fig.~\ref{fig:views} shows the results of the RPC-PET
reconstruction applied to Geant4 simulated data based on the NCAT
anthropomorphic phantom.
The images on the left of Fig.~\ref{fig:views}
represent the coronal and sagittal views of the NCAT simulation
phantom. The images on the right of the same figure show
the reconstruction images of 49~M true events after 20 MLEM
iterations. A lower cut on the intensity scale 
and on the following reconstructed images was performed in order to remove the
very-low intensity from voxels filled with the tails of the Gaussian
TOF kernel.

Axial, coronal and sagittal views of the brain are also included
in Figs.~\ref{fig:trues_axial_views} and~\ref{fig:brain} demonstrating the capabilities of RPC-PET to reveal the detailed
structures of the brain.

\begin{figure}[H]
\centering
\includegraphics[width=3.3in]{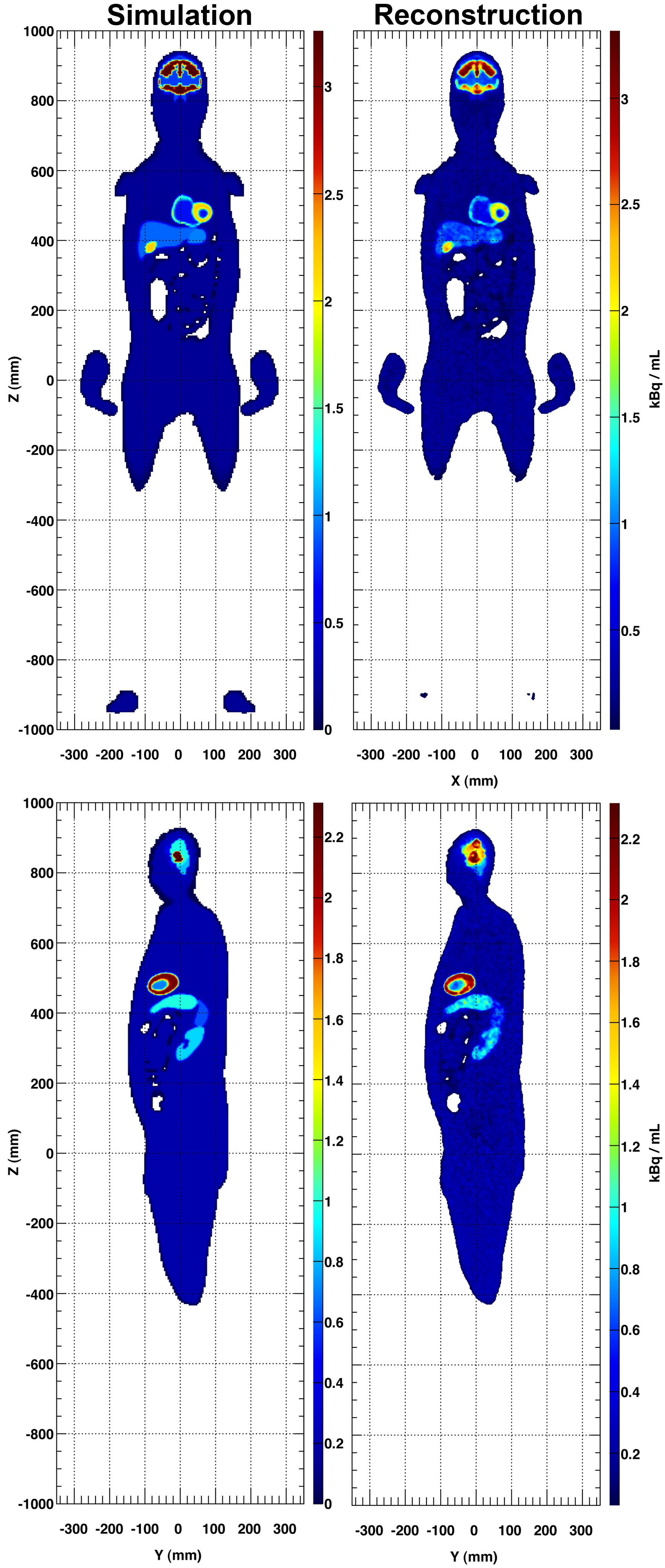}
\vspace{-2mm}
\caption{Results of the RPC-PET reconstruction applied to
  Geant4 simulated data based on the NCAT anthropomorphic phantom
  (left column). The right column shows reconstructed results after 20
  MLEM iterations. The two views represent 2 mm thick
  slices across the heart region. The sagittal view shows the left
  ventricle, stomach, spleen and kidney, while the coronal view shows both ventricles
  of the heart well separated from the liver and stomach. Gallbladder and
  intensity depressions on the intestine region are also visible.
}
\label{fig:views}
\end{figure}

\begin{figure}[H]
\centering
\includegraphics[width=3.2in]{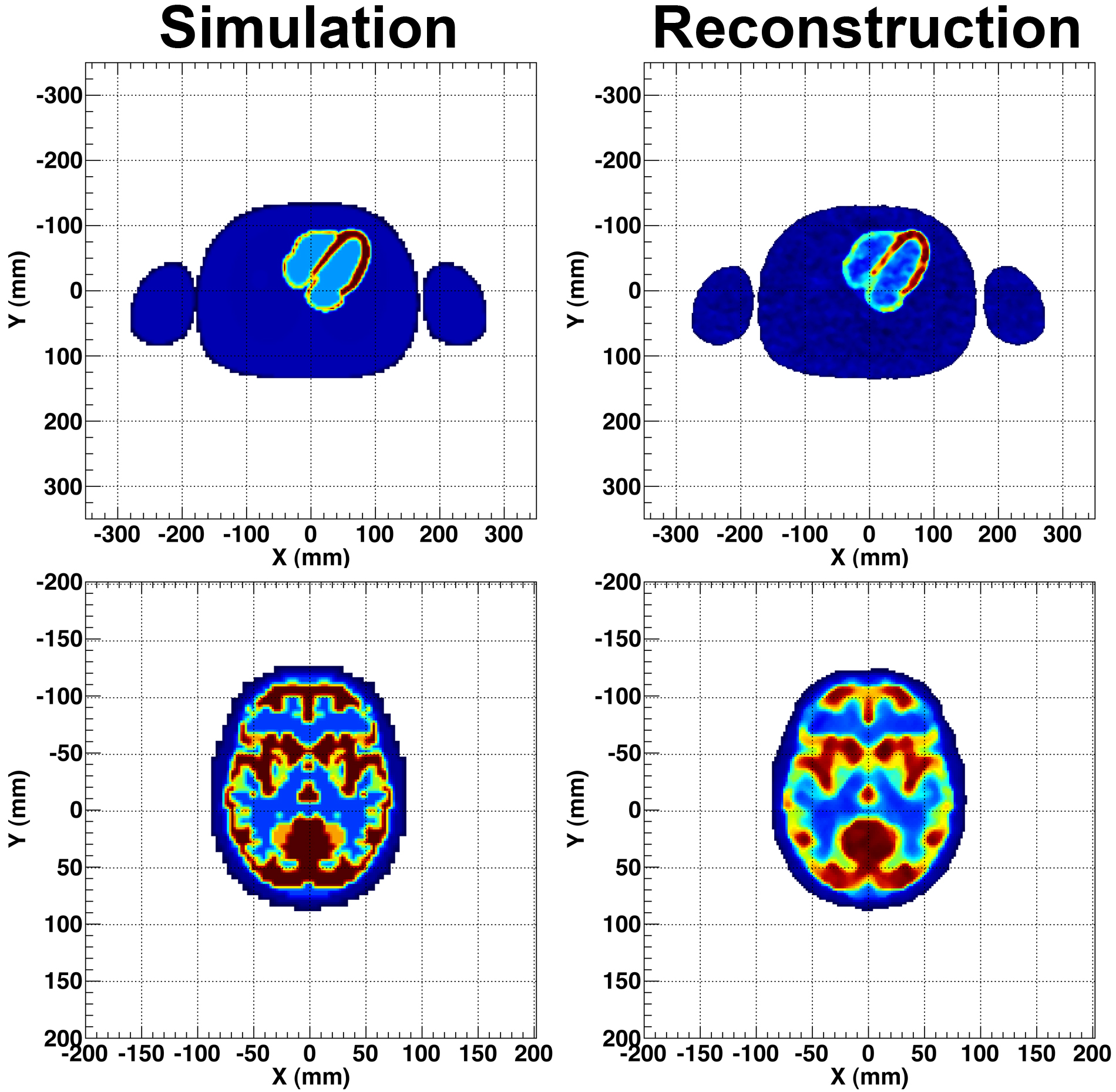}
\vspace{-3mm}.
\caption{Two-millimeter-thick axial slices of the brain and the
  heart. This reconstruction offers very high resolution images of the
  heart walls and the brain regions simultaneously.}
\label{fig:trues_axial_views}
\end{figure}

\begin{figure}[H]
\centering
\includegraphics[width=3.25in]{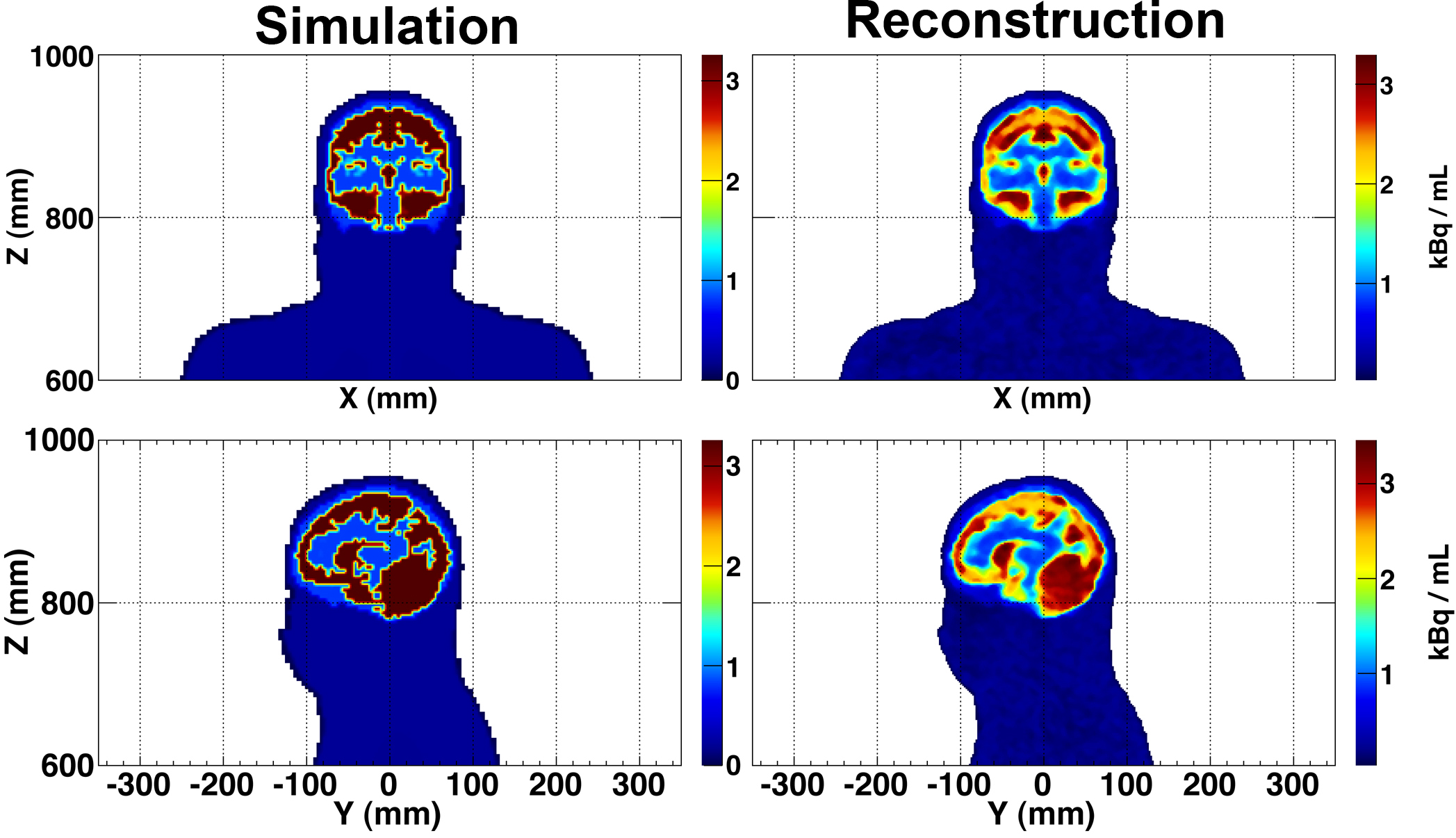}
\vspace{-4mm}
\caption{Coronal and axial views of the brain. A very good agreement between the cerebellum and
  other activated regions of the brain is observed, revealing detailed structural
  information of the brain.}
\label{fig:brain}
\end{figure}

\subsection{Scatter Rejection Fraction}

Fig.~\ref{fig:SF} shows that the lack of energy resolution in the RPC-PET
detectors is compensated for by their very-high TOF resolution. Indeed,
by applying the scatter rejection method described in
section~\ref{sec:SRF_method} (Fig.~\ref{fig:SRF_draw}), one sees in
Fig.~\ref{fig:SF} a SF of 32.9\% for 300~ps FWHM, which compares with
a SF of 57.1\% if no scatter rejection by TOF was applied. The scatter
rejection fraction (SRF), i.e., the ratio between the number of scattered
events rejected with TOF and the total of scattered events detected, was
63\% for a 300~ps FWHM TOF resolution. More than half of the
scattered LORs did not cross the NCAT phantom and were therefore also
rejected by this method, even for a 4~ns TOF resolution.

\begin{figure}[H]
\centering
\includegraphics*[width=7.75cm]{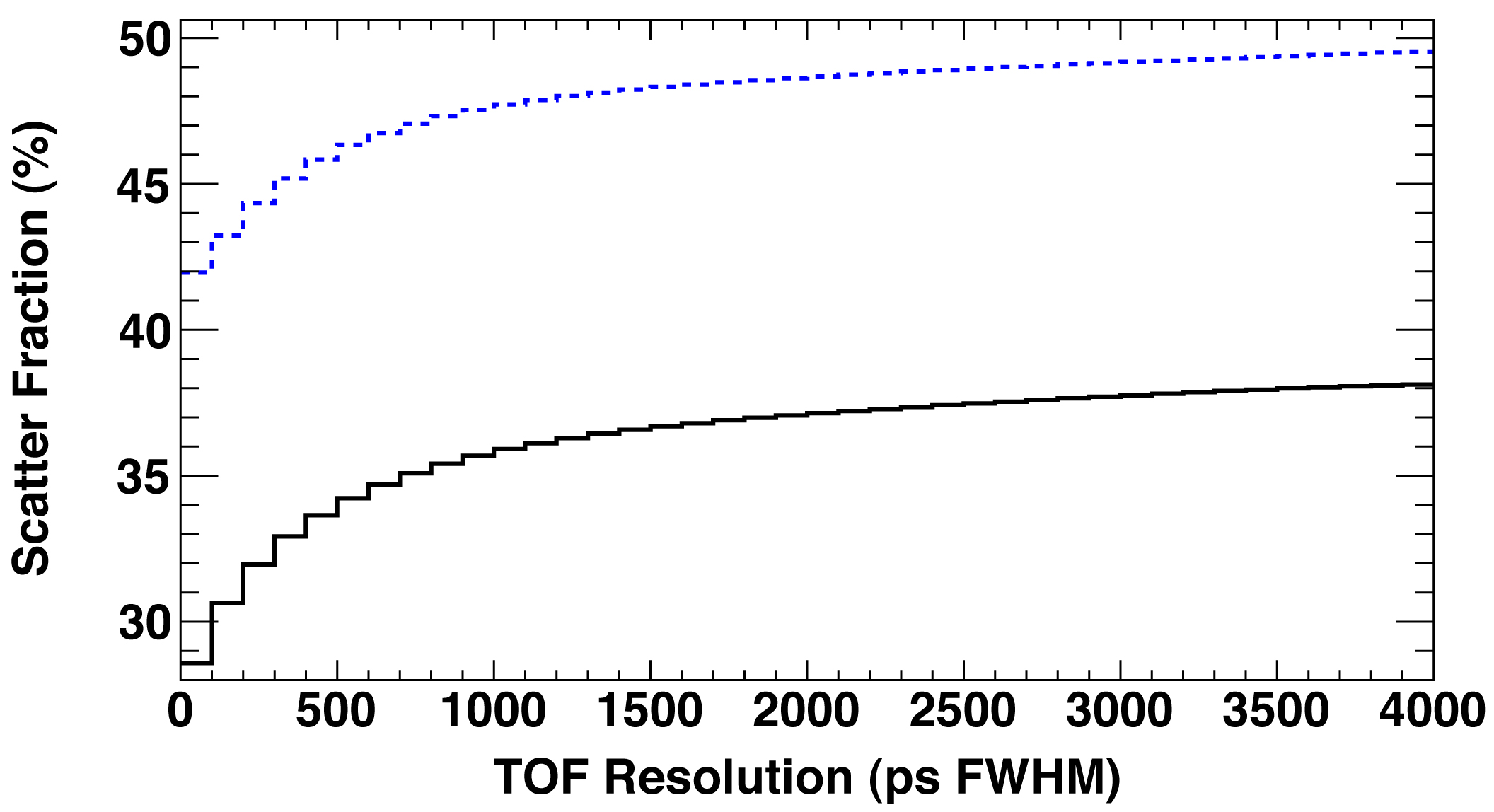}
\vspace{-2mm}
\caption{Curve of the body scatter fraction (SF) for a decreasing TOF
  resolution, calculated for the homogeneous cylindrical phantom
  (dashed blue) and for the NCAT phantom (solid black).
For a 300 ps FWHM TOF resolution, the SF is 32.9\% and 45.2\% for the
homogeneous cylindrical phantom and the NCAT phantom,
respectively. For a 600 ps FWHM TOF resolution, these values increase
to 34.7\% and 46.7\%, respectively. }
\label{fig:SF}
\end{figure}

In the case of the spheres immersed in a water phantom, the SF
decreased from 67.2\% (no TOF) to 45.2\% with a 300~ps FWHM TOF-based scatter
rejection.

\subsection{Six lesions in an anthropomorphic phantom: towards lesion detectability}

\begin{figure}[!t]
\includegraphics[width=.47\textwidth]{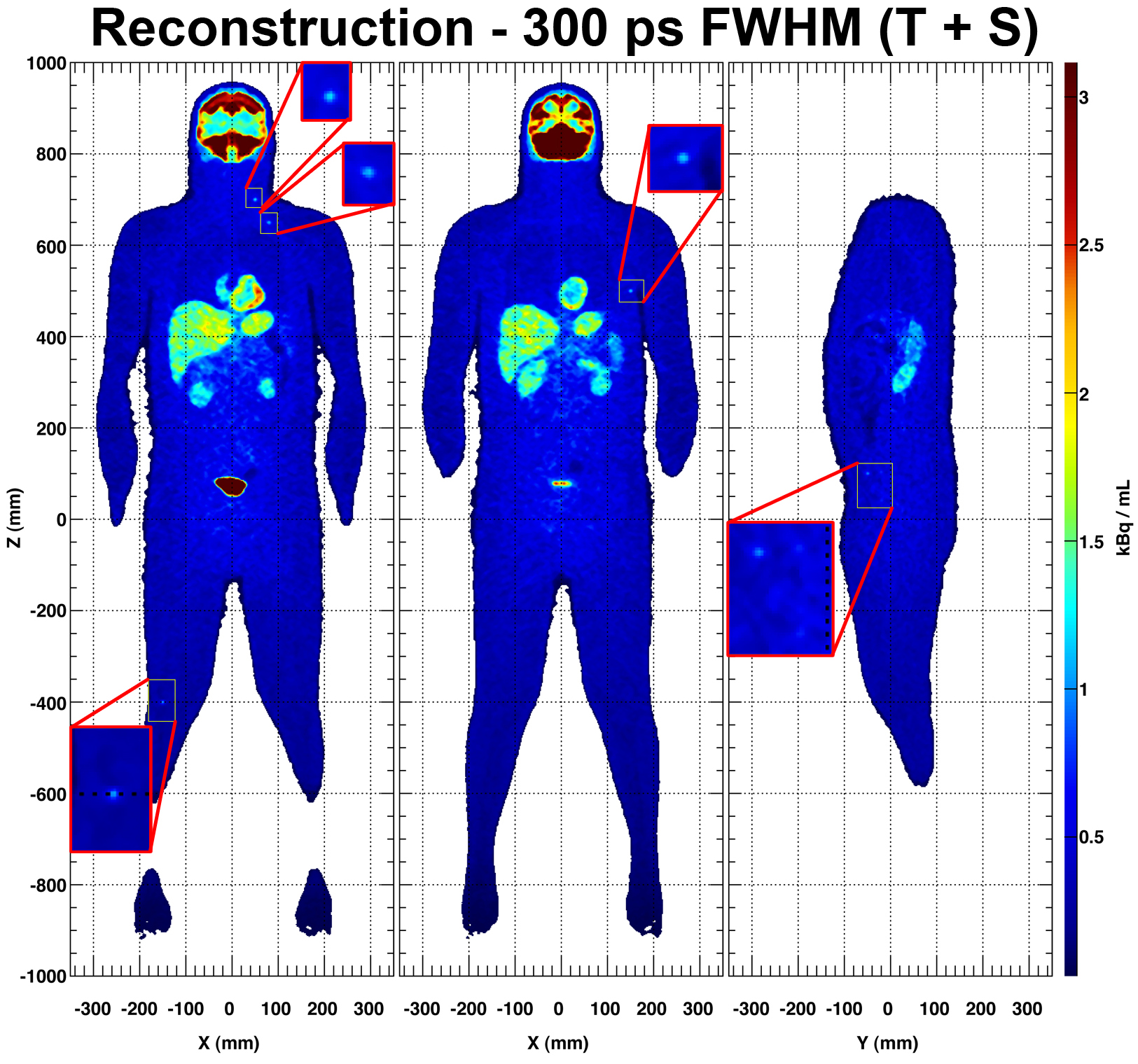}
\hfill
\includegraphics[width=.47\textwidth]{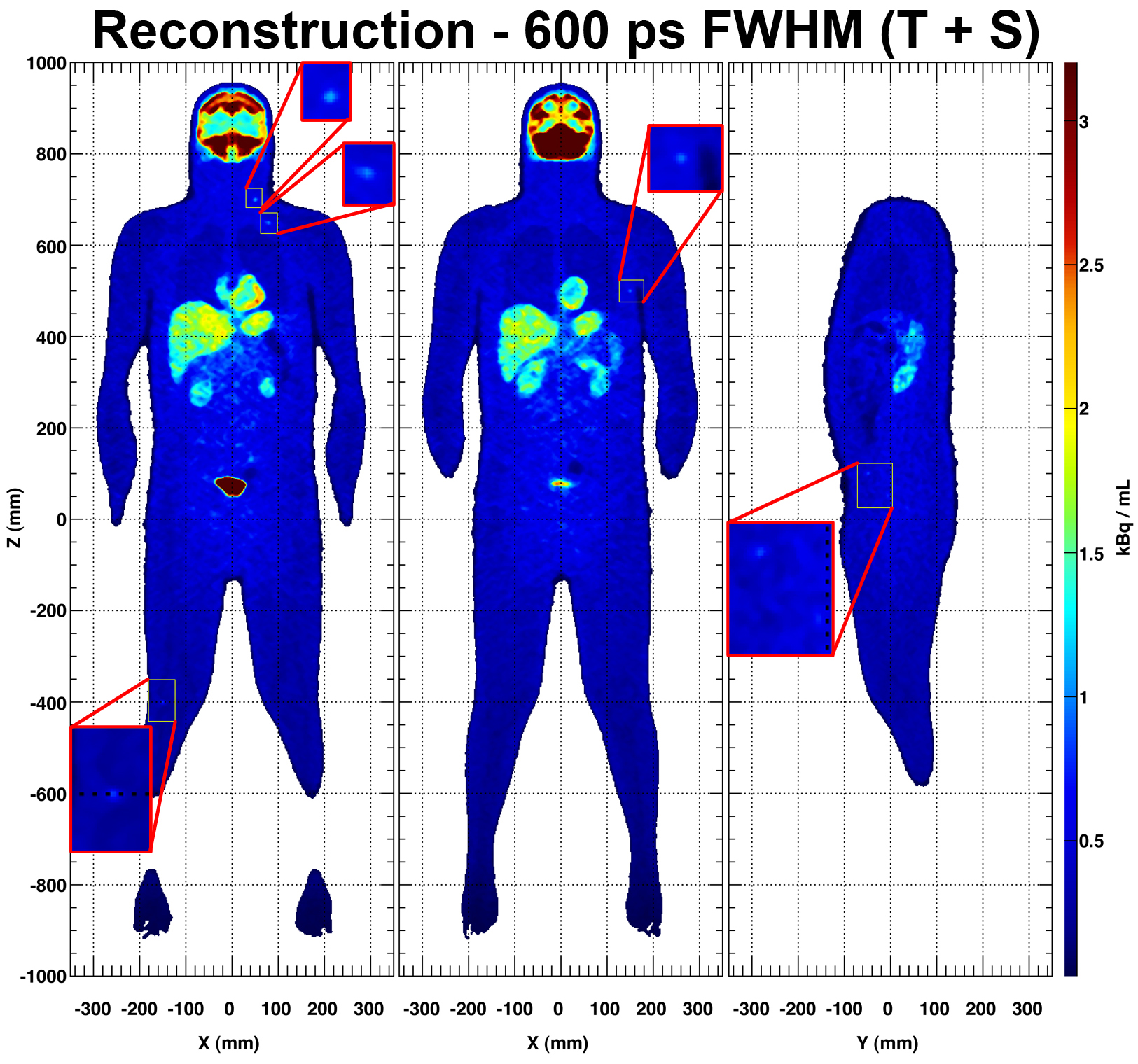}

\vspace{-4mm}
\caption{Whole-body 2-mm-thick slice images showing five lesions:
  cervical, sub-clavicular, knee, axillary, and
  inguinal. The reconstructed results were obtained from the data
  (trues and scatters) detected by an RPC-PET system for a 300~ps (top) and
a 600~ps (bottom) FWHM TOF kernel. It is a well-known fact that the abdominal region is more
contaminated by scatters; however, this region becomes cleaner due to
the scatter rejection properties of RPC-PET. The organs are still well
separated: the heart from the stomach and the right kidney from the
liver. The spleen is still visible. The visual contrast improves in all
lesions for a 300~ps FWHM TOF resolution (top), primarily the inguinal one,
which is almost indistinguishable
with a 600~ps FWHM TOF resolution (bottom). Images were obtained after
20 MLEM iterations. A windowing on the intensity scale was performed to
  distinguish the lesions.}
\label{fig:rpc_SR}
\end{figure}

Fig.~\ref{fig:rpc_SR} shows two sets of 2-mm-thick slice images,
each with five simulated lesions. 
All figures hereinafter presented containing the three views of
the five lesions were
normalized to 80\% of the maximum intensity in the brain.
In Fig.~\ref{fig:rpc_SR},
we included the true and body-scattered events detected by an
RPC-PET system with 120 gaps after passing by an energy selection
curve~\cite{Crespo2012}. We then compared the 300~ps and
600~ps FWHM TOF kernels after performing the TOF-based scatter
rejection (see Fig.~\ref{fig:SRF_draw}). The comparison with the 600~ps FWHM TOF case was in line
with the state-of-the-art commercial PET scanners.

Compared to the reconstructed images after performing the
rejection cuts with a 600~ps FWHM TOF kernel
(Fig.~\ref{fig:rpc_SR}, bottom), all the lesions observed in
the 300~ps FWHM images (Fig.~\ref{fig:rpc_SR}, top) had improved
visual contrast, primarily the inguinal lesion, which was almost indistinguishable using the
600~ps FWHM TOF kernel. The contribution
of the scatters in the abdomen region was highly suppressed.
Therefore, the combination of
an RPC-PET detector with a higher TOF resolution increases
visual detectability, mainly in regions more sensitive to the contribution of
scattered events, such as those closer to the abdomen.
A more comprehensive study of the constrast recovery for each lesion will
be partly presented in the following sections.

To better distinguish each of the six lesions we show the corresponding axial
slices crossing the maximum intensity voxel of each lesion in
Fig.~\ref{fig:6lesions}. The line profiles across the lesion are presented in
Fig.~\ref{fig:6lesions} (bottom). They discriminate the presence
of a lesion in all six cases, including the inguinal one, whose visual detection
was not fully evident in the coronal and sagittal views in
Fig.~\ref{fig:rpc_SR}. This is due to its location in the outer region
of the abdomen, where contaminating background is less prominent than in
the center. 

\begin{figure}[H]
\centering
\includegraphics[width=.45\textwidth]{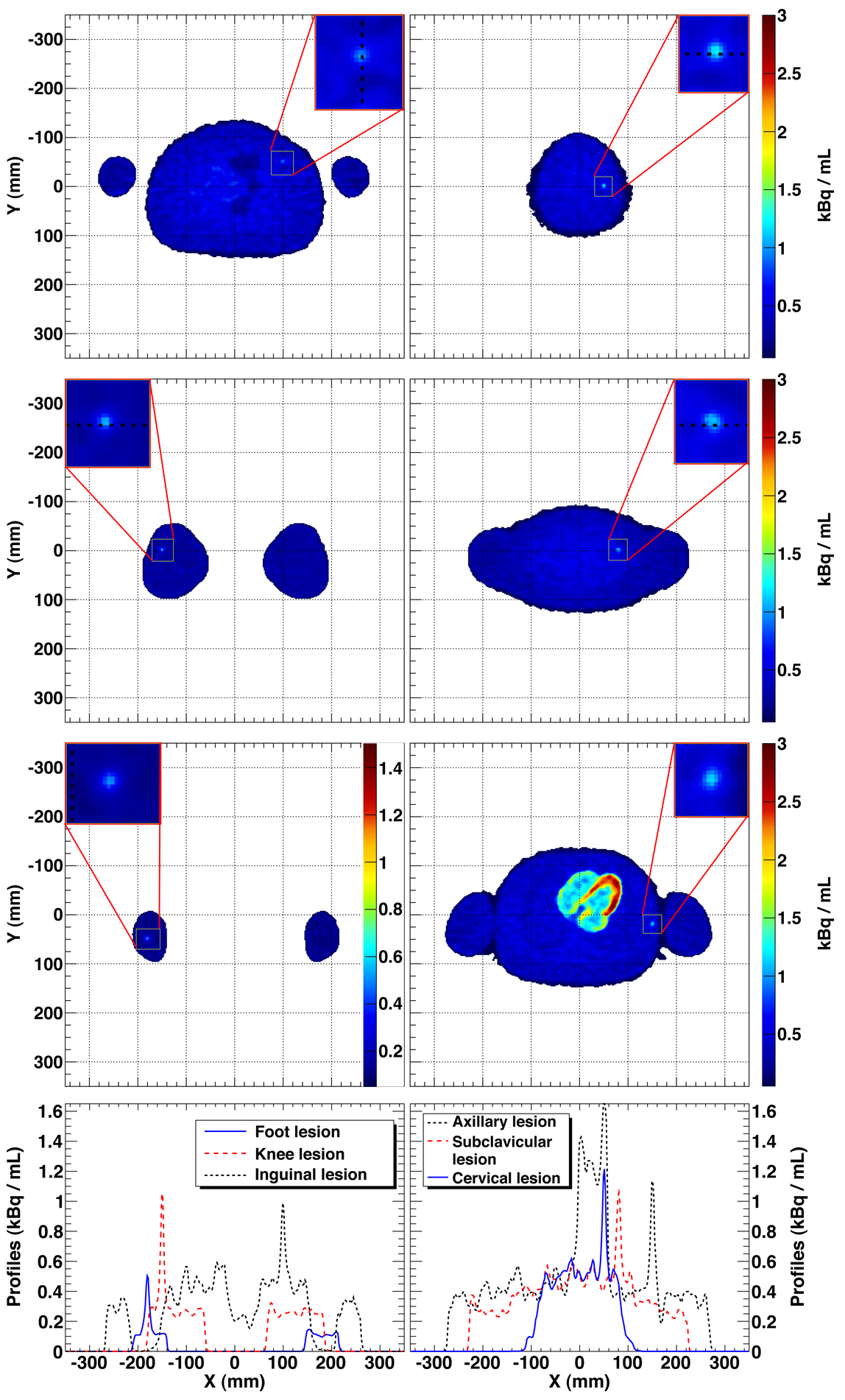}
\vspace{-4mm}
\caption{Axial views corresponding to 2 mm thick slices crossing the
  six lesions. In the left column are shown the foot
  (bottom); knee (middle); and inguinal (top) lesions. In the
  right column are shown the axillary (bottom); subclavicular
  (middle); and cervical (top) lesions. The profile across the inguinal
lesion (dashed black curve) clearly indicates the
presence of a lesion.
} 
\label{fig:6lesions}
\end{figure}

The knee lesion had the
best visual contrast, while the axillary lesion appeared with good
visual contrast
despite the nearby heart within the image. All images,
except the one containing the foot lesion, were normalized to the
maximum activity in the heart. The foot lesion lost activity due to
its presence in the extremity of the RPC detection system (not covered by the whole solid angle acceptance) and was then normalized
to half of the maximum intensity in the heart walls.

\subsection{Performance}
\subsubsection{RMSE Evolution}

Fig.~\ref{fig:RMSE} presents the RMSE evolution for the MLEM reconstruction
along 200 iterations for the NCAT (top), and along 80 iterations for the
spheres in a homogenous background (bottom). 

\begin{figure}[H]
\centering
\includegraphics*[width=3.2in]{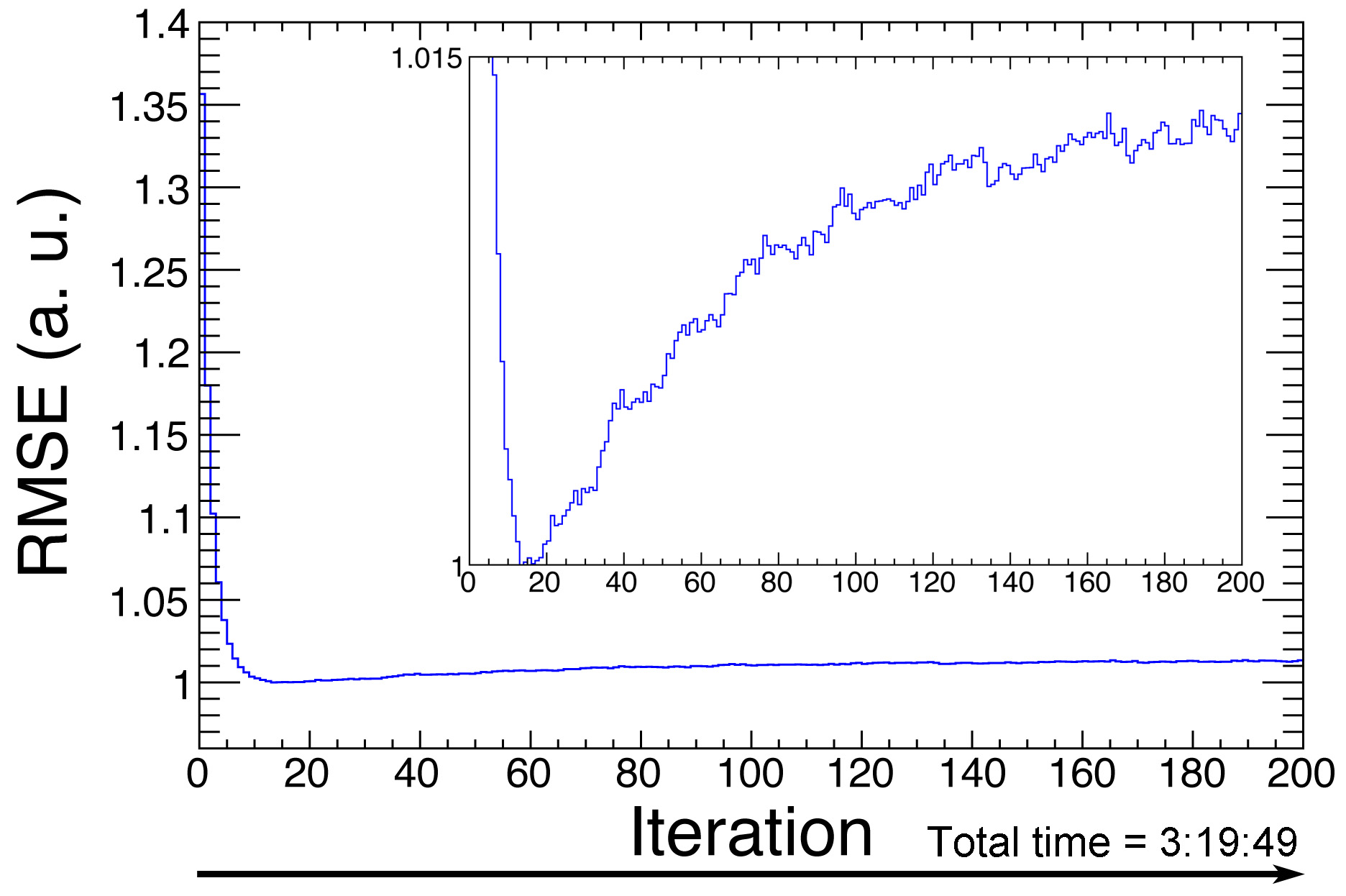}
\vfill
\includegraphics*[width=3.2in]{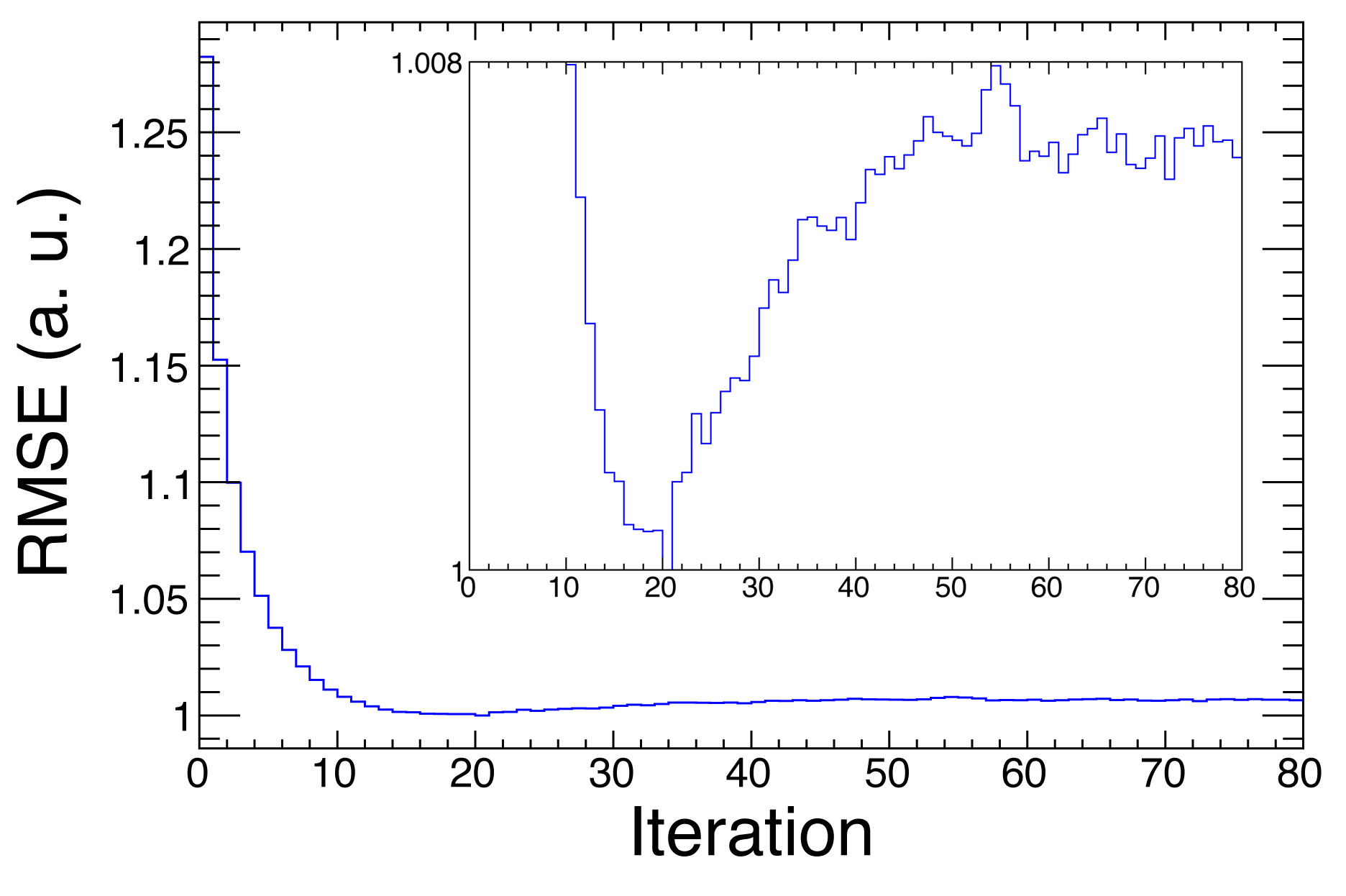}
\vspace{-3mm}
\caption{Graph of the evolution of the RMSE along 200 iterations
  calculated between the NCAT simulated image and the
  reconstructed true events image (top) and along 80 iterations
  calculated between the mathematical image of the spheres in a
  homogeneous background and the reconstructed true events image
  (bottom). In both, the minimum is reached after 20 iterations.}
\label{fig:RMSE}
\end{figure}

The RMSE was calculated
between the simulated (NCAT) or mathematical image (water cylinder) and the reconstructed true
events image. Reconstruction with scattered events was not
considered, due to a consequent bias, since scatter correction was not
performed. The RMSE reached a minimum after 20 iterations, a
fact that led us to the decision of stopping
the reconstruction at iteration 20 (in this study).

\subsubsection{Image Division}

\begin{figure}[H]
\centering
\includegraphics[width=3.5in]{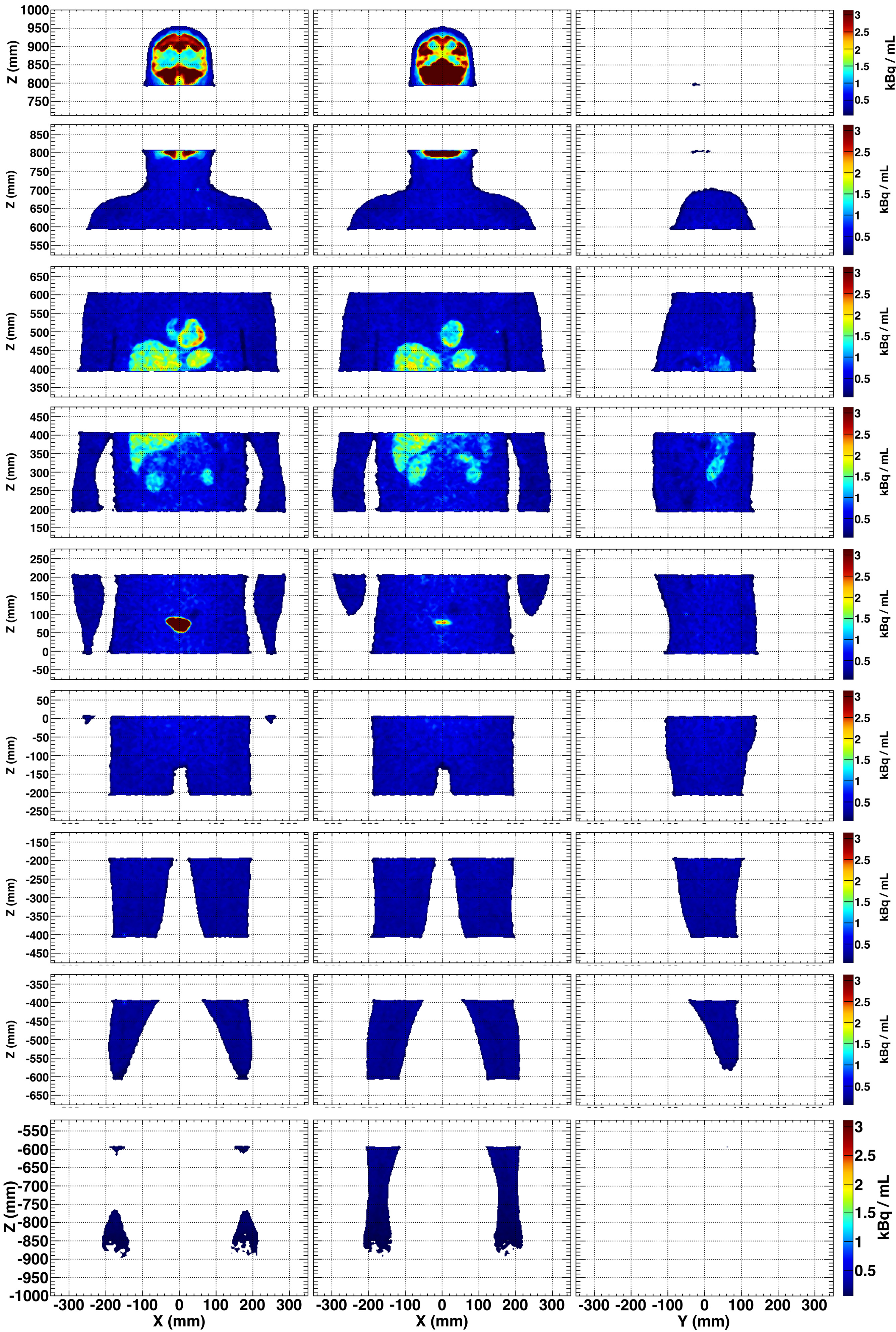}
\caption{Reconstruction method based on the division of the data
  through nine different regions of the body. This image 
  represents the nine independent reconstructions, obtained after 20 MLEM
  iterations. This strategy allows for the full whole-body
  reconstruction to be performed in only 3.5 minutes, which compares
  with 21 minutes if the whole body is considered.
}
\label{fig:division_1}
\end{figure}

As an alternative approach to a whole-body reconstruction, we
divided the reconstructed image based on the data arising
from nine different regions of the body. Fig.~\ref{fig:division_1}
presents the results of the nine independent reconstructions after 20
MLEM iterations, whose resulting images were then summed and are
shown in Fig.~\ref{fig:division_2}.
This method improved
the reconstruction speed as presented later in table~\ref{tb:speed_table}. A
margin of 3~$\sigma$ (as explained in section~\ref{sec:data_division_method}) had to be given in order to consider the events
laying outside the image while iterating on the TOF-kernel.

A median filter tends to reduce the intensity in a voxel if several 
sur\-round\-ing voxels are null, which occurs in the
borders of those divided images. Therefore, an outer shell of 6~mm (3
voxels in both $Z$-directions) was added and an overlap between
neighbouring images was performed.

\begin{figure}[H]
\centering
\includegraphics[width=3.5in]{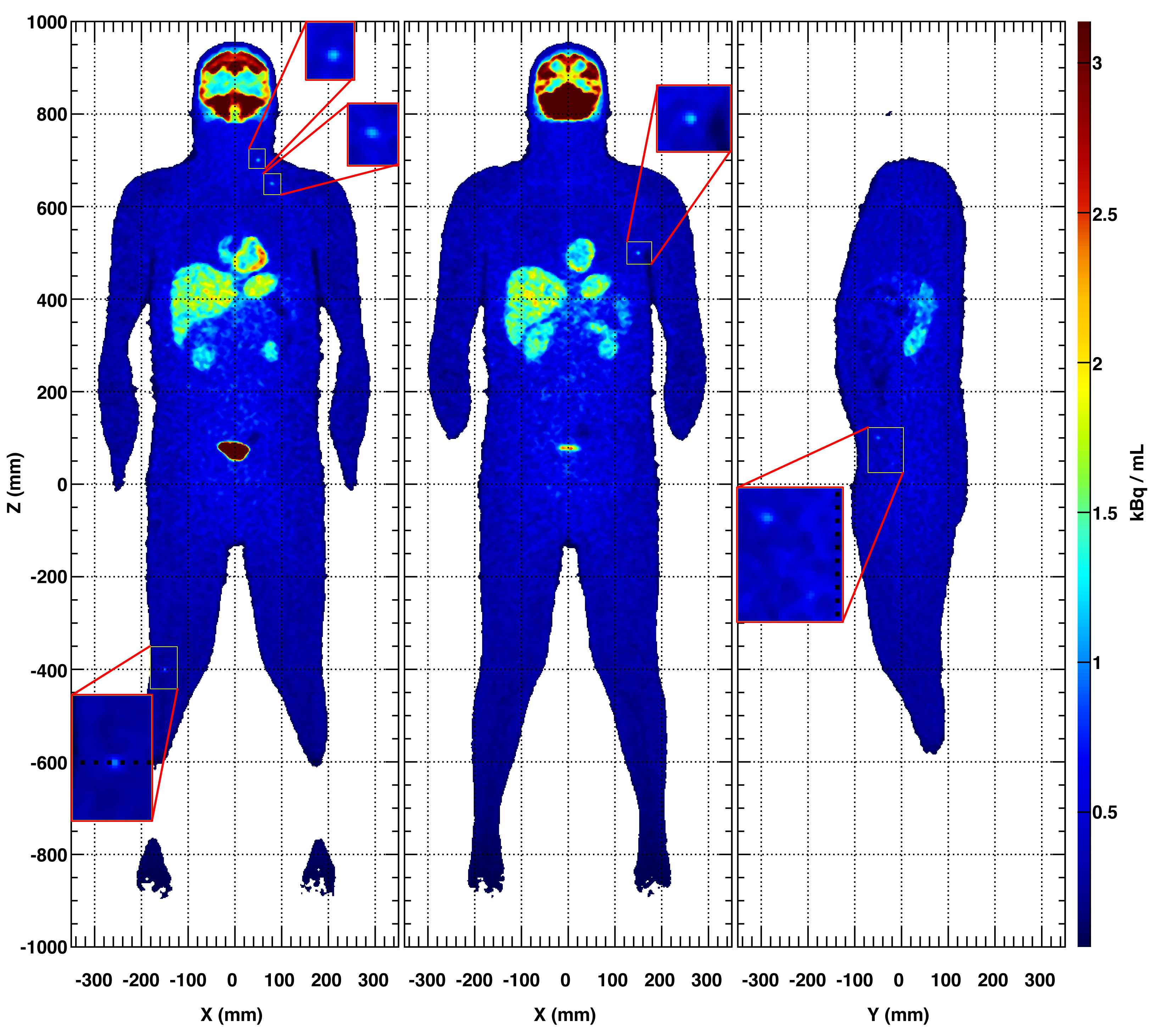}
\vspace{-7mm}
\caption{
  Reconstructed image resulting from the sum of the nine 
  independent reconstructions depicted in Fig.~\ref{fig:division_1}. There is no
  difference to the whole-body reconstruction shown in
  Fig.~\ref{fig:rpc_SR} (top).
}
\label{fig:division_2}
\end{figure}

As depicted in
Fig~\ref{fig:division_3} (left), the knee lesion positioned in the
divided image borders is not affected by this method and its profile
(bottom left) compares well with the one in Fig.~\ref{fig:6lesions}. 

This image division method also allowed a quick
detection of un\-ex\-pect\-ed lesions, like the one in the foot presented in
Fig.~\ref{fig:division_3} (right). Even for this region, in which the
RPC-PET detector was less sen\-si\-tive, the lesion was clearly
visible, mainly due to the ab\-sence of other surrounding organs with higher
uptake values.

\begin{figure}[H]
\centering
\includegraphics[width=3.5in]{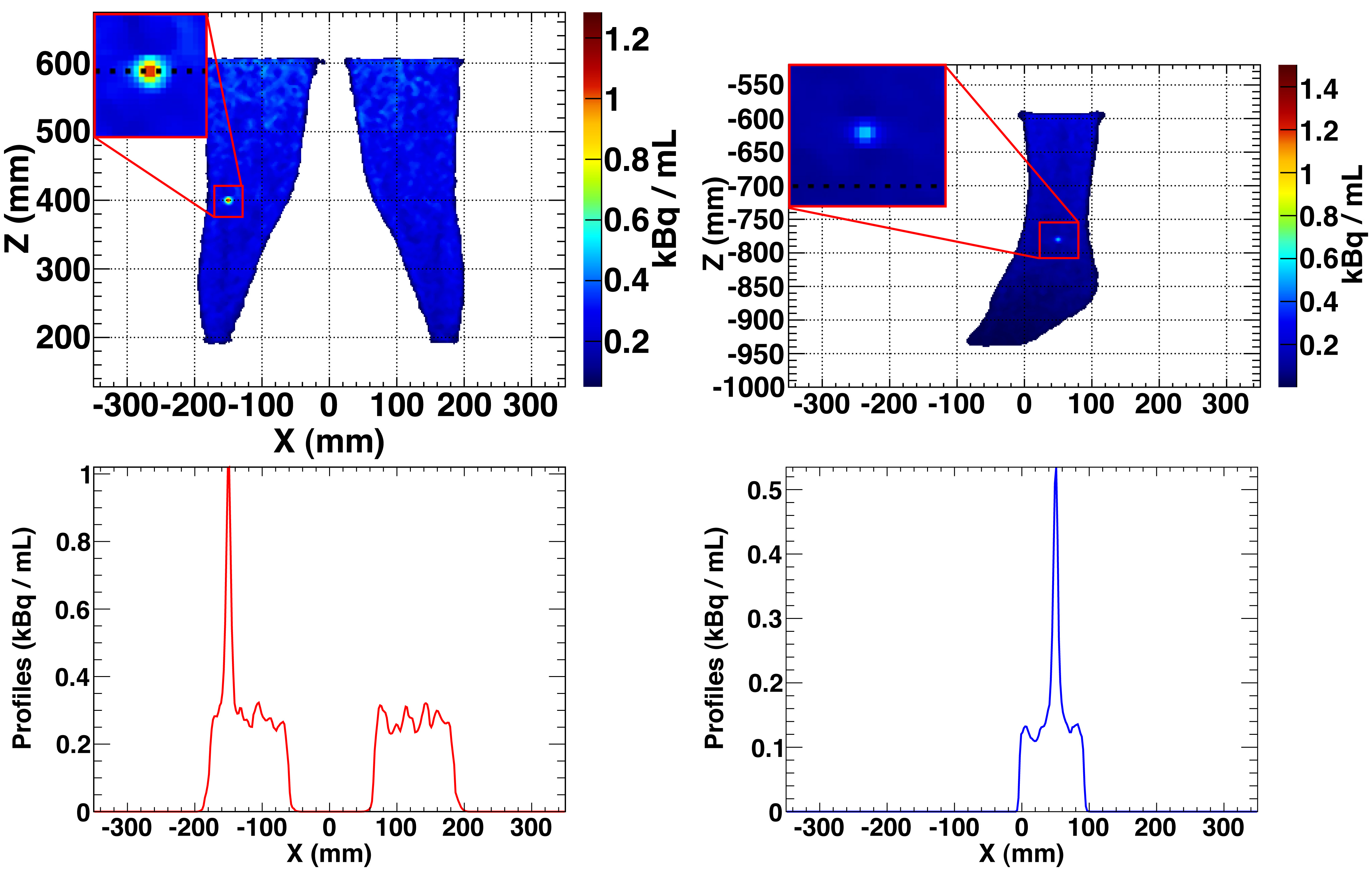}
\vspace{-7mm}
\caption{Reconstructed images of the legs (left) and the right foot
  (right). The activity in the inner part of the
  leg where the two images sum is not affected and the lesion is well
  visible. Performing an analysis of each of the nine independent
  reconstructions, it is possible to detect a lesion in
  the foot that could go undetected in a whole-body reconstruction
  due to the lack of intensity in this region with respect to the rest
  of the body, the latter being fully covered by the solid angle of
  the RPC-PET system.
}
\label{fig:division_3}
\end{figure}

\subsubsection{Contrast recovery coefficients}

The CRC values for the six spheres in the homogeneous phantom obtained
from the reconstruction of the true events, with a TOF resolution of
300~ps FWHM, and at iteration 20 (MLEM) are presented in table~\ref{tb:table_CRC_spheres}.

\begingroup
\begin{table}[hbtp]
\tiny
\centering
\renewcommand{\arraystretch}{1.3}
\renewcommand{\tabcolsep}{0.1cm}
 \caption[CRC values for six spheres in
   homogeneous phantom.]{CRC values for six spheres in
   homogeneous phantom (\%)}
\begin{tabular}{@{}lcccccc@{}}
\toprule
& \multicolumn{6}{c}{\head{Trues (300~ps)}} \\
\cline{2-7}
Spheres & 1 & 2 & 3 & 4 & 5 & 6\\
\cline{1-7}
Iter. 20 & 18.1 $\pm$ 2.0 & 19.4 $\pm$ 2.1  &  24.3 $\pm$ 3.6 &17.7 $\pm$ 2.0 &  22.0 $\pm$ 2.4 &  16.0 $\pm$ 1.4 \\
\end{tabular}
\label{tb:table_CRC_spheres}
\end{table}
\endgroup

We verified that the CRC values for the lesions in the NCAT phantom
obtained with the GPU and the 16-thread CPU implementations of the
reconstruction routines were quite similar (not shown). 

As seen in Fig.~\ref{fig:CRC_evolution}, there was a
large gain between iterations 1 and 20 and a stabilization with an
increasing number of iterations. 
\begin{figure}[!t]
\centering
\includegraphics[width=.49\textwidth]{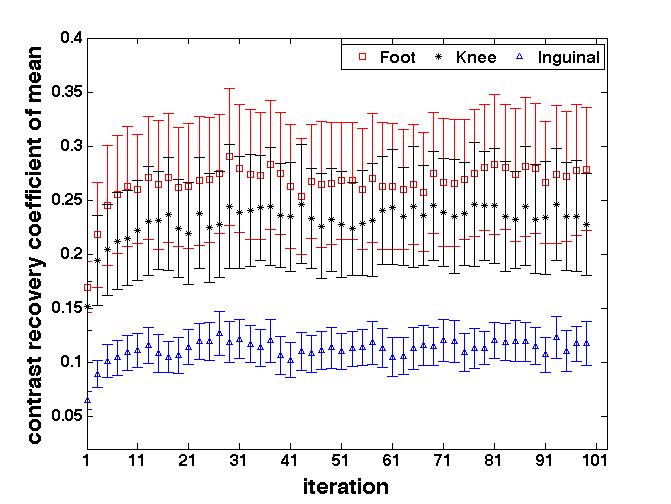}
\vfill
\includegraphics[width=.49\textwidth]{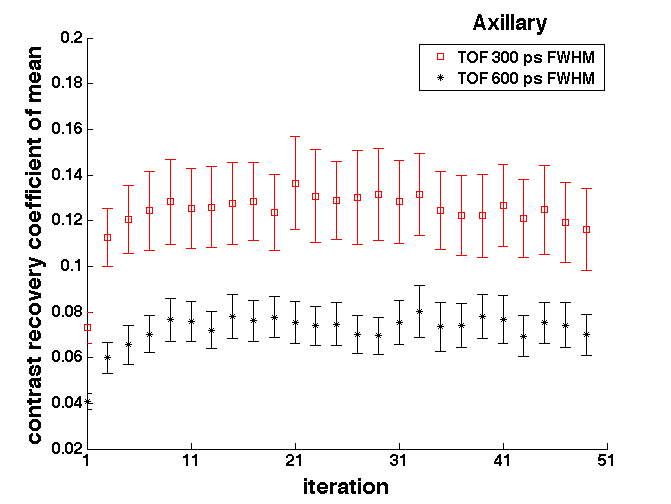}
\caption{Contrast recovery coefficient evolution with
    iterations for the images reconstructed with true and scatter
    events (MLEM, SRF (300~ps FWHM)). Top: CRC in foot,
    knee and inguinal lesions. Bottom: comparison between a TOF resolution of 300
    and 600~ps FWHM for the axillary 
    lesion.
}
\label{fig:CRC_evolution}
\end{figure}

We present the CRC 
in the foot, knee and inguinal
lesions (top). The CRC took into account the region average
for each lesion. The CRC values in other lesions evolved similarly,
i.e., they stabilized after iteration 20 (not shown).
The advantage of a TOF resolution of 300~ps FWHM is clear for the
axillary lesion (bottom).

\subsubsection{Speed}

For the pre-processing and the starting backprojection that feeds the
reconstruction routines, we selected 16-thread CPUs
alone because the time gain for the GPU implementation of those routines
was not remarkable. 

The reconstruction times presented
in table~\ref{tb:speed_table} already
took into account the contribution of the pre-processing and backprojection times.
Both reconstruction routines ran faster with the GPU plus 16-thread CPU
implementation than with the 16-thread CPUs alone. 

Table~\ref{tb:speed_table}
also presents the reconstruction time for the presented images, taking
into account
the two different approaches: the whole-body and division methods. 
For the whole-body, GPU
took 21:11 to perform the reconstruction of the data for the 
SRF (300~ps FWHM). 

Considering the image division approach, a
performance time below 4 minutes (3:33, 20 iterations) for the
superior torso region was accomplished -- the worst case in terms of
computational burden. This time included the total image
reconstruction process.

\vspace{-5mm}

\section{Discussion}

The large AFOV and the 300~ps FWHM TOF resolution were combined in the
RPC-PET to make this detector relevant for low injection doses and
short scan periods. We expect to obtain the images presented in this work
with a 7~minute scan and an injected activity of 2~mCi, a value well
below the activities currently administered to patients (10 to 20~mCi). 

A reconstruction routine
capable of providing images for RPC-PET technology was
created. A direct TOF implementation of the MLEM
algorithms allows for all events to be directly processed and
inserted inside the object image by means of a TOF kernel, while
handling list-mode data iteratively. A proper attenuation correction
was successfully implemented. 

A method of rejection of the scattered events was presented, leading
to a reduction of the body scatter fraction from 57.1\% to 32.9\% for a 300~ps FWHM TOF
whole-body RPC-PET system.

To study the potential of TOF resolution on lesion detectability, we
evaluated its impact on the detection of six simulated lesions
spread over critical regions within the anthropomorphic phantom. Lesions
close to the abdomen, such as the
inguinal lesion, are harder to detect. In comparison with a
600~ps FWHM TOF detector (current commercial crystal-based PET scanners), the 300~ps FWHM whole-body
RPC-PET may be a more sensitive detector, representing a potential
increase in detectability.

As an alternative approach to whole-body reconstruction, a method based on the division
of the data through nine different regions was proposed,
resulting in an enhancement in the reconstruction performance. The proposed
method provided a reconstruction that was six times faster. Making use of
GPUs assisted by 16-thread CPUs, we expect to reach a reconstructed
image from a 300~ps FWHM RPC-PET scanner in 3.5 minutes after the end of data
acquisition. Compared to our first attempts to reconstruct
whole-body data in a single CPU, which took approximately 90 hours,
this represents a significant progress.

\begin{table}[!t]
\tiny
\renewcommand{\arraystretch}{1.3}
\renewcommand{\tabcolsep}{0.2cm}
 \caption{Reconstruction time (minutes) for the
presented images. Timing in bold indicates largest computing time.}

\begin{tabular}{@{}llccccc@{}}
\toprule
\multicolumn{1}{c}{}
&\multicolumn{1}{c}{}
 & \multicolumn{4}{c}{\head{SRF (300~ps) - Body divided into nine regions}} \\
\cmidrule(r){3-6}
&Body Region &Head & Head & Torso & Torso \\
&& Superior&Inferior & Superior&Inferior \\
\cmidrule(lr){1-6}
\parbox[t]{2mm}{\multirow{4}{*}{\rotatebox[origin=c]{90}{Pre-process.}}}
\vspace{0.2cm}
&Scatter Rejection & 0:06 & 0:02 & 0:05 & 0:04\\
&Attenuation correction  & 0:06 & 0:03 & 0:05 & 0:04 \\
\cmidrule(lr){2-6}
&Total  & 0:12 & 0:05 & 0:10 & 0:08 \\
\cmidrule(lr){1-6}
&Backprojection & 0:14 & 0:11 & 0:13 & 0:13 \\
\cmidrule(lr){1-6}
\parbox[t]{2mm}{\multirow{4}{*}{\rotatebox[origin=c]{90}{MLEM 20 it.}}}
\vspace{0.1cm}
&16-thread CPUs& 4:13 & 3:38 & 4:44 & 4:28 \\
\\
&GPU (Tesla C2075)  & 3:30 & 2:48 & {\bf{3:33}} & 3:14 \\
\\
\end{tabular}

\renewcommand{\tabcolsep}{0.1cm}
\begin{tabular}{llccccc}
\toprule
\multicolumn{1}{c}{}
&\multicolumn{1}{c}{}
 & \multicolumn{5}{c}{\head{SRF (300~ps) - Body divided into nine regions}} \\
\cmidrule(r){3-7}
&Body Region &Abdominal & Abdominal & Legs & Legs \\
&& Superior&Inferior & Superior&Inferior & Feet\\
\cmidrule(lr){1-7}
\parbox[t]{2mm}{\multirow{4}{*}{\rotatebox[origin=c]{90}{Pre-process.}}}
\vspace{0.2cm}
&Scatter Rejection & 0:03 & 0:03 & 0:03 & 0:02 & 0:02\\
&Attenuation correction  & 0:04 & 0:03 & 0:03 & 0:02 & 0:01 \\
\cmidrule(lr){2-7}
&Total  & 0:07 & 0:06 & 0:06 & 0:04 & 0:03 \\
\cmidrule(lr){1-7}
&Backprojection & 0:13 & 0:11 & 0:10 & 0:10 & 0:12 \\
\cmidrule(lr){1-7}
\parbox[t]{2mm}{\multirow{4}{*}{\rotatebox[origin=c]{90}{MLEM 20 it.}}}
\vspace{0.2cm}
&16-thread CPUs & 4:20 & 4:06 & 3:51 & 3:32 & 3:51\\
\\
&GPU (Tesla C2075) & 3:04 & 2:51 & 2:36 & 2:22 & 2:40 \\
\\
\bottomrule
\end{tabular}

\renewcommand{\tabcolsep}{0.2cm}
\begin{tabular}{llccccc}
\multicolumn{1}{c}{}
&\multicolumn{1}{c}{}
 & \multicolumn{4}{c}{\head{Whole Body}} \\
\cmidrule(r){3-6}
&&SRF &SRF &Trues  & Trues\\
&& (300~ps)&(600ps) & (300~ps)&(600~ps) \\
\cmidrule(lr){1-6}
\parbox[t]{2mm}{\multirow{4}{*}{\rotatebox[origin=c]{90}{Pre-process.}}}
\vspace{0.2cm}
&Scatter Rejection & 0:27 & 0:44 & - & -\\
&Attenuation correction  & 0:41 & 0:52 & 0:25 & 0:25 \\
\cmidrule(lr){2-6}
&Total  & 1:08 & 1:36 & 0:25 & 0:25 \\
\cmidrule(lr){1-6}
&Backprojection & 1:40 & 2:16 & 1:20 & 1:53 \\
\cmidrule(lr){1-6}
\parbox[t]{2mm}{\multirow{4}{*}{\rotatebox[origin=c]{90}{MLEM 20 it.}}}
\vspace{0.2cm}
&16-thread CPUs & 33:52 & - & 27:00 & - \\
\\
&GPU (Tesla C2075)  & {\bf 21:11} & - & 17:03 & - \\
\\
\bottomrule
\end{tabular}

\label{tb:speed_table}
\end{table}

A CRC analysis corroborates the conclusions we have drawn thus far: the CRC
stabilizes at iteration 20; the
inner lesions in the body have a higher loss of contrast due to the
contribution of the scattered events in those regions; the TOF
resolution dictates the largest contribution to contrast gain.

\section{Conclusion}

A reconstruction routine capable of providing images for single-bed
whole-body RPC-PET
technology was demonstrated. An enhancement of reconstruction speed
was provided, as well as a method for body-scatter rejection. Future
work includes appropriate random and scatter corrections and experimental
phantom studies.

\vspace{-3mm}

\section*{Acknowledgments}
The authors greatly acknowledge support from the High Performance
Computing Center of the University of Coimbra, Portugal, from
Dr. Miguel Oliveira former LIP member, from the LIP helpdesk, Gon\c{c}alo
Borges and Jo\~ao Martins, and from Jo\~ao
Silva (LIP). The authors also thank Prof. Dr. Paul Segars, from Johns
Hopkins University, for providing the software-based anthropomorphic
phantom.
This work was supported by the EU, FEDER, POCI, QREN, COMPETE, POFC,
PORC, MaisCentro and by the Portuguese Government through Foundation
for Science and Technology (FCT) and Comiss\~ao de Coordena\c{c}\~ao e
Desenvolvimento Regional do
Centro (CCDRC), under the contracts CERN/\-FP/\-123605/\-2011,
PTDC/\-SAU-BEB/\-104630/\-2008 and
CENTRO\--07\--ST24\--FEDER\--002007 (project Rad4Life), co-funded by the European Social Fund and
by POPH - Programa Operacional Potencial Humano. 
P.~Crespo, M.~Couceiro, and	P.~Martins acknowledge FCT grants
numbers SFRH/BPD/39223/2007, SFRH/BD/42217/2007, and
SFRH/BPD/103655/2014, respectively.

\vspace{-3mm}

\section*{Disclosure statement}

The authors report no conflicts of interest. The authors alone are responsible
for the content and writing of this article.

\vspace{-3mm}

\bibliographystyle{tfnlm}
\bibliography{arXiv_RPCPET_Rec}

\begin{thebibliography}{10}
\providecommand{\url}[1]{\normalfont{#1}}
\providecommand{\urlprefix}{Available from: }

\bibitem{FonteNIMA2000}
Fonte~P, Smirnitsky~A, Williams~MCS. A new high-resolution {TOF} technology.
  Nucl Instrum Meth A. 2000;\hspace{0pt}443(1):201--204.

\bibitem{BlancoNIMA2003}
Blanco~A, Chepel~V, {Ferreira Marques}~R, et~al. Perspectives for positron
  emission tomography with {RPC}s. Nucl Instrum Meth A.
  2003;\hspace{0pt}508:88--92.

\bibitem{Martins2014}
Martins~P, Blanco~A, Crespo~P, et~al. Towards very high resolution rpc-pet for
  small animals. Journal of Instrumentation. 2014;\hspace{0pt}9(10):C10012.

\bibitem{BlancoTNS2006}
Blanco~A, Carolino~N, Correia~CMBA, et~al. {RPC--PET}: A new very high
  resolution {PET} technology. IEEE Trans Nucl Sci.
  2006;\hspace{0pt}53(5):2489--2494.

\bibitem{Blanco2009}
Blanco~A, Couceiro~M, Crespo~P, et~al. Efficiency of {RPC} detectors for
  whole-body human {TOF--PET}. Nucl Instrum Meth A.
  2009;\hspace{0pt}602(3):780--783.

\bibitem{Werner2006}
Werner~ME, Surti~S, Karp~JS. Implementation and evaluation of a {3D PET} single
  scatter simulation with {TOF} modeling. In: Conf. Record of the 2006 IEEE
  Nucl. Sci. Symp. \& Med. Imag. Conf. (NSS/MIC); Vol.~3; Oct; San Diego, CA,
  USA; 2006. p. 1768--1773.

\bibitem{Watson2007}
Watson~CC. Extension of single scatter simulation to scatter correction of
  time-of-flight {PET}. IEEE Trans Nucl Sci. 2007
  Oct;\hspace{0pt}54(5):1679--1686.

\bibitem{CouceiroNIMA2012}
Couceiro~M, Crespo~P, Mendes~L, et~al. Spatial resolution of human {RPC--PET}
  system. Nucl Instrum Meth A. 2012;\hspace{0pt}661, Supplement 1(0):S156 --
  S158.

\bibitem{Crosetto2003}
Crosetto~DB. {The 3D complete body screening (3D-CBS) features and
  implementation}. In: Conf. Record of the 2003 IEEE Nucl. Sci. Symp. \& Med.
  Imag. Conf. (NSS/MIC); Vol.~4; Oct; Portland, OR, USA; 2003. p. 2415--2419.

\bibitem{Eriksson2008}
Eriksson~L, Townsend~DW, Conti~M, et~al. Potentials for large axial field of
  view positron camera systems. In: Conf. Records 2008 IEEE Nucl. Sci. Symp. \&
  Med. Imag. Conf. (NSS/MIC); Oct. 19--25; Dresden, Germany; 2008. p.
  1632--1636.

\bibitem{Borasi2010}
Borasi~G, Fioroni~F, Del~Guerra~A, et~al. {PET} systems: the value of added
  length. Eur J Nucl Med Mol Imaging. 2010;\hspace{0pt}37(9):1629--1632.

\bibitem{Poon2012}
Poon~JK, Dahlbom~ML, Moses~WW, et~al. Optimal whole-body {PET} scanner
  configurations for different volumes of {LSO} scintillator: a simulation
  study. Phys Med Biol. 2012;\hspace{0pt}57(13):4077.

\bibitem{Shepp1982}
Shepp~L, Vardi~Y. Maximum likelihood reconstruction for emission tomography.
  IEEE Trans Med Imaging. 1982 Oct;\hspace{0pt}1(2):113--122.

\bibitem{Groiselle2004}
Groiselle~CJ, Glick~SJ. {3D PET} list-mode iterative reconstruction using
  time-of-flight information. In: Conf. Record of the 2004 IEEE Nucl. Sci.
  Symp. \& Med. Imag. Conf. (NSS/MIC); Vol.~4; Oct; Rome, Italy; 2004. p.
  2633--2638.

\bibitem{karpJNM2008}
Karp~JS, Surti~S, Daube-Witherspoon~ME, et~al. Benefit of time-of-flight in
  {PET}: experimental and clinical results. J Nucl Med.
  2008;\hspace{0pt}49:462--470.

\bibitem{Crespo2007}
Crespo~P, Shakirin~G, Fiedler~F, et~al. Direct time-of-flight for quantitative,
  real-time in-beam {PET}: a concept and feasibility study. Phys Med Biol.
  2007;\hspace{0pt}52:6795--–6811.

\bibitem{Witherspoon2012}
Daube-Witherspoon~ME, Matej~S, Werner~ME, et~al. {Comparison of list-mode and
  DIRECT approaches for time-of-flight PET reconstruction}. IEEE Trans Med
  Imag. 2012 July;\hspace{0pt}31(7):1461--1471.

\bibitem{Conti2013}
Conti~M, Eriksson~L, Westerwoudt~V. Estimating image quality for future
  generations of {TOF PET} scanners. IEEE Trans Nucl Sci. 2013
  Feb;\hspace{0pt}60(1):87--94.

\bibitem{Pratx2009}
Pratx~G, Chinn~G, Olcott~PD, et~al. Fast, accurate and shift-varying line
  projections for iterative reconstruction using the {GPU}. IEEE Trans Med
  Imaging. 2009 March;\hspace{0pt}28(3):435--445.

\bibitem{Pratx2011}
Pratx~G, Surti~S, Levin~CS. Fast list-mode reconstruction for time-of-flight
  {PET} using graphics hardware. IEEE Trans Nucl Sci. 2011
  Feb;\hspace{0pt}58(1):105--109.

\bibitem{Cui2011}
Cui~J, Pratx~G, Prevrhal~S, et~al. Fully {3D} list-mode time-of-flight {PET}
  image reconstruction on {GPUs} using {CUDA}. Medical Physics.
  2011;\hspace{0pt}38(12):6775--6786.

\bibitem{Sportelli2013}
Sportelli~G, Ortu\~no~JE, Vaquero~JJ, et~al. Massively parallelizable list-mode
  reconstruction using a {Monte Carlo-based} elliptical {Gaussian} model.
  Medical Physics. 2013;\hspace{0pt}40(1):012504.

\bibitem{Segars2001}
Segars~WP. Development of a new dynamic {NURBS}-based cardiac-torso ({NCAT})
  phantom [dissertation]. The University of North Carolina, USA; 2001.

\bibitem{Crespo2012}
Crespo~P, Reis~J, Couceiro~M, et~al. Whole-body single-bed time-of-flight
  {RPC--PET}: simulation of axial and planar sensitivities with {NEMA} and
  anthropomorphic phantoms. IEEE Trans Nucl Sci.
  2012;\hspace{0pt}59(3):520--529.

\bibitem{Paquet2004}
Paquet~N, Albert~A, Foidart~J, et~al. {Within-patient variability of 18F-FDG:
  standardized uptake values in normal tissues}. Journal of Nuclear Medicine.
  2004;\hspace{0pt}45(5):784--788.

\bibitem{Swanson1990}
Swanson~DP, Chilton~HM, Thrall~JH. {Pharmaceuticals in medical imaging}. New
  York: Macmillan Publishing Company; 1990.

\bibitem{Zhang2007}
Zhang~J, Olcott~PD, Chinn~G, et~al. {Study of the performance of a novel 1 mm
  resolution dual-panel PET camera design dedicated to breast cancer imaging
  using Monte Carlo simulation}. Medical Physics.
  2007;\hspace{0pt}34(2):689--702.

\bibitem{Hwu2010}
Kirk~DB, Hwu~WW. Programming massively parallel processors. Massachussets:
  Morgan Kaufmann; 2010.

\bibitem{Sanders2011}
Sanders~J, Kandrot~E. Cuda by example. Massachussets: Addison Wesley; 2011.

\bibitem{Martins2014b}
Martins~PJM. {Imaging Techniques in RPC-PET} [dissertation]. University of
  Coimbra, Portugal; 2014.

\bibitem{Hebert1990}
Hebert~TJ, Leahy~R. {Fast methods for including attenuation in the EM
  algorithm}. IEEE Trans Nucl Sci. 1990 Apr;\hspace{0pt}37(2):754--758.

\bibitem{Comtat1998}
Comtat~C, Kinahan~PE, Defrise~M, et~al. Fast reconstruction of {3D PET} data
  with accurate statistical modeling. IEEE Trans Nucl Sci. 1998
  Jun;\hspace{0pt}45(3):1083--1089.

\bibitem{Levkovitz2001}
Levkovitz~R, Falikman~D, Zibulevsky~M, et~al. {The design and implementation of
  COSEM, an iterative algorithm for fully 3-D listmode data}. IEEE Trans Med
  Imag. 2001 July;\hspace{0pt}20(7):633--642.

\bibitem{Snyder1985}
Snyder~DL, Miller~M. The use of sieves to stabilize images produced with the em
  algorithm for emission tomography. IEEE Trans Nucl Sci. 1985
  Oct;\hspace{0pt}32(5):3864--3872.

\bibitem{Tong2010}
Tong~S, Alessio~AM, Kinahan~PE. {Noise and signal properties in PSF-based fully
  {3D PET} image reconstruction: an experimental evaluation}. Phys Med Biol.
  2010;\hspace{0pt}55(5):1453.

\end{thebibliography}

\end{twocolumn}

\end{document}